\documentclass[11pt,a4paper,reqno]{amsart}
\usepackage[T1]{fontenc}
\usepackage[utf8]{inputenc}
\usepackage[english]{babel}
\usepackage{lmodern,microtype}
\usepackage{amssymb}
\usepackage[margin=2.6cm]{geometry}
\usepackage[hidelinks]{hyperref}
\usepackage{xcolor}
\usepackage{tikz}
\usetikzlibrary{arrows.meta,calc,shapes.misc}

\numberwithin{equation}{section}

\theoremstyle{plain}
\newtheorem{theorem}{Theorem}[section]
\newtheorem{lemma}[theorem]{Lemma}
\newtheorem{proposition}[theorem]{Proposition}
\newtheorem{corollary}[theorem]{Corollary}

\theoremstyle{definition}
\newtheorem{definition}[theorem]{Definition}

\theoremstyle{remark}
\newtheorem{remark}[theorem]{Remark}

\newcommand{\R}{\mathbb R}
\newcommand{\tr}{\operatorname{tr}}
\newcommand{\Id}{\mathbf 1_2}
\newcommand{\dd}{\,\mathrm d}

\title[Dropping Three Bodies]{Dropping Three Bodies:\\Moment of Inertia and Side Ordering}

\author{Moritz Link}
\address{Faculty of Physics and Astronomy, Friedrich Schiller University Jena}
\email{moritz.link@uni-jena.de}

\author{Lorenz Osburg}
\address{Faculty of Physics and Astronomy, Friedrich Schiller University Jena}
\email{lorenz.osburg@uni-jena.de}

\author{Moritz Scholz}
\address{Faculty of Physics and Astronomy, Friedrich Schiller University Jena}
\email{moritz.scholz@uni-jena.de}

\author{Manuel Quaschner}
\address{Faculty of Mathematics and Computer Science, Friedrich Schiller University Jena}
\email{manuel.robert.quaschner@uni-jena.de}

\date{30 November 2026}

\begin{document}
	
\setlength{\parindent}{0pt}	

	\begin{abstract}
		We study the planar three-body problem for solutions released from rest (brake orbits) at a noncollinear configuration, and prove two statements formulated for Newtonian gravity as conjectures by \textsc{Moeckel}, \textsc{Montgomery} and \textsc{Venturelli}. First, for arbitrary positive masses and symmetric strictly attractive central pair forces, the first syzygy is reached in finite time and the moment of inertia is strictly decreasing until then, with $\dot I_m < 0$ also at the syzygy itself if it is regular. Second, for equal masses and $0 < \alpha < 2$, a scalene initial triangle reaches a collision-free first syzygy, and the strict ordering of its side lengths is preserved up to and including this syzygy. In particular, the body opposite the initially longest side is the middle body at the first syzygy. The first result follows from a matrix \textsc{Riccati} equation for the \textsc{Jacobi} configuration matrix, in the spirit of \textsc{Tsygvintsev}. The second rests on a torque equation for the rotation of each side and a kinematic identity linking side differences, area and rotation, together with an energy estimate that excludes binary collisions for $\alpha < 2$.
	\end{abstract}

	\maketitle
	
	\section{Introduction}
	\subsection{$n$-body problem}
	We want to study the motion of three bodies started from rest. For this, we first introduce the general framework of the $n$-body problem and introduce some notation. \\
	The $n$-body problem is the study of the dynamics of $n$ point particles with masses $m_i>0$ and positions $q_i \in \mathbb{R}^d$ according to \textsc{Newton}'s second law
	\begin{subequations}\label{eq:equations_of_motion}
\begin{equation}\label{eq:alpha_forces}
    m_i\ddot q_i
    =
    \alpha G
    \sum_{\substack{\ell=1\\\ell\neq i}}^n
    \frac{m_i m_\ell}
    {|q_\ell-q_i|^{\alpha+2}}
    (q_\ell-q_i),
    \qquad 1\leq i\leq n.
\end{equation}
	Here $G > 0$ is the gravitational constant and $\alpha > 0$ is a given parameter defining a generalised $\alpha$-potential. For standard gravitational interactions, we take $\alpha = 1$. \\
	More generally, we consider symmetric, strictly attractive
	central pair forces, described by
	\begin{equation}\label{eq:general_central_forces}
		m_i\ddot q_i
		=
		\sum_{\substack{\ell=1\\\ell\neq i}}^n
		\kappa_{i\ell}(|q_\ell-q_i|)
		(q_\ell-q_i),
		\qquad 1\leq i\leq n,
	\end{equation}
	\end{subequations}
	where the interaction coefficients satisfy
	\[
		\kappa_{i\ell}=\kappa_{\ell i},
		\qquad
		\kappa_{i\ell}\in
		C^1\bigl((0,\infty),(0,\infty)\bigr).
	\]
	In particular, the $\alpha$-potential equations
	\eqref{eq:alpha_forces} are recovered by choosing
	\[
		\kappa_{i\ell}(r)
		=
		\frac{\alpha Gm_i m_\ell}{r^{\alpha+2}}.
	\]
	Thus, \eqref{eq:general_central_forces} includes the
	$\alpha$-potentials as a special case without requiring
	homogeneity of the interaction. 
	Our first main result
	will hold for this more general class of attractive forces,
	whereas the second result concerns the $\alpha$-potential
	setting.

Let $q=(q_1, \dots ,q_n)\in \mathbb{R}^{nd}$ be the \emph{configuration vector} and write
$r_{ij}=|q_j-q_i|$ for the Euclidean distance. Since the forces in
\eqref{eq:general_central_forces} are central and depend only on the distance, they are
conservative: for each pair choose a pair potential
$\Phi_{ij}=\Phi_{ji}\in C^2\bigl((0,\infty)\bigr)$ with
\begin{equation}\label{eq:pair_potential}
	\Phi_{ij}'(r) = -r\,\kappa_{ij}(r) \quad r>0,
\end{equation}
which is unique up to an additive constant, and define the \emph{potential}
\begin{equation}\label{eq:general_potential}
	U(q)=\sum_{i<j}\Phi_{ij}(r_{ij}).
\end{equation}
Since $\nabla_i r_{ij} = (q_i-q_j)/r_{ij}$, we have
$\nabla_i \Phi_{ij}(r_{ij}) = \kappa_{ij}(r_{ij})\,(q_j-q_i)$, and hence
\eqref{eq:general_central_forces} takes the form
\begin{equation}
	m_i\ddot{q}_i = \nabla_i U(q), \quad 1 \leq i \leq n,
\end{equation}
where $\nabla_i$ denotes the $d$-dimensional partial gradient with respect to $q_i$.
For the $\alpha$-potential we choose $\Phi_{ij}(r) = G m_i m_j r^{-\alpha}$, which gives
the generalised \emph{Newtonian potential}
\begin{equation}\label{eq:newton_potential}
	U(q)=G\sum_{i<j}\frac{m_im_j}{r_{ij}^\alpha}.
\end{equation}
In matrix notation, one can write
\begin{equation}
	\mathbf{M}\ddot{q} = \nabla U(q),
\end{equation}
where $\nabla$ denotes the $nd$-dimensional gradient and
$\mathbf{M} = \operatorname{diag}(m_1, \dots, m_n)$ is the diagonal mass matrix with $d$ copies of
each mass along the diagonal (see, e.g., \cite{Moe14}). Since the coefficients
$\kappa_{ij}$ are only defined at positive separations, and the $\alpha$-potential is
singular at collisions, the \emph{configuration space} is the $nd$-dimensional manifold
$\mathcal{M}=\mathbb{R}^{nd}\setminus \Delta$, where
\begin{equation}
	\Delta = \bigcup_{1 \le i < j \le n} \Delta_{ij},
	\qquad
	\Delta_{ij} = \{ q \in \mathbb{R}^{nd} \mid q_i = q_j \},
\end{equation}
is the \emph{singular set} \cite[Chapter~10]{AM}. Introducing the velocity variable
$v=\dot q$, Newton's equations can be rewritten as the first-order system
\[
\dot q=v,\qquad \dot v=\mathbf{M}^{-1}\nabla U(q)
\]
on the phase space $\mathcal P=\mathcal M\times\mathbb R^{nd}$, see \cite{Moe14}. Since
$\kappa_{ij}\in C^1$, the vector field is continuously differentiable on $\mathcal P$, and
real analytic for the $\alpha$-potential. Thus, the \textsc{Picard}--\textsc{Lindelöf}
theorem guarantees a unique maximal solution for every initial condition
$(q_0,v_0)\in\mathcal P$.\\
	For Newtonian gravitation, unlike the superintegrable two-body problem, the dynamics of the system for $n>2$ are generally not integrable in the sense of \textsc{Liouville}--\textsc{Arnold} theory. As is well-known, \textsc{Bruns} \cite{Bruns1887} (compare also \cite[Section~11.3.3]{Kn18}) showed in 1887 that the $n$-body problem admits no independent integrals of motion that are algebraic functions on the phase space, besides the ten classical integrals (total linear momentum, total angular momentum, center of mass, and energy). For this reason, mathematicians focus on qualitative properties of solutions rather than searching for explicit solutions to \eqref{eq:alpha_forces}.\\

\subsection{Main results}	
In this paper, we consider a problem in the planar three-body problem, setting $d=2$ and $n=3$. Dropping the three bodies from rest corresponds to the initial velocity condition $\dot{q}_i(0)=0$ for $i \in \{1,2,3\}$, so that the dynamics is governed solely by the interaction forces. A solution having an instant at which all velocities vanish is called a \emph{brake orbit}. For the $\alpha$-potential model, brake orbits have negative energy and zero angular momentum. The vanishing of angular momentum also holds for general rotational symmetric central forces by Noether's theorem. A \emph{syzygy} is a collinear configuration (see \cite{MMV} for these definitions). We define $T_{*}$ as the endpoint of the initial noncollinear interval. If the solution reaches a regular syzygy at this endpoint, it is the time of the first syzygy after our start from rest. Throughout this paper, a \emph{regular} syzygy means that the three positions are distinct. In \cite[Theorem~1.1]{MontSyzygy}, \textsc{Montgomery} counts binary collisions as syzygies and excludes the \textsc{Lagrange} solution, a brake orbit that ends in a triple collision (for further information, see \cite{MMV}). In this sense, \eqref{eq:general_central_forces} is well-defined at regular syzygies. Montgomery in  \cite{MontSyzygy} also proves that $T_*$ is finite for standard Newtonian gravity. \\
We prove the following two statements—originally formulated for Newtonian gravity as Conjectures~1 and~2 in \cite{MMV}—for general attractive central forces in the first case and for $\alpha$-potentials in the second, with solutions starting from rest at a noncollinear configuration. All system variables are defined in Section~\ref{sc:preliminaries}.
 \\
The first statement investigates the moment of inertia $I_m$. The \textsc{Lagrange}-\textsc{Jacobi}-identity shows for a brake orbit that the moment of inertia has to decrease initially. However, it is not clear how long this has to be the case. We prove that the moment of inertia decreases strictly monotone at least until the time of the first syzygy. The mathematical precise formulation is as follows:

\begin{theorem}\label{thm:inertia}
	For arbitrary $m_1,m_2,m_3>0$ and symmetric, strictly attractive coefficients $\kappa_{ij}\in C^1((0,\infty),(0,\infty))$, one has $T_* < \infty$ and the solution of
	\eqref{eq:general_central_forces} starting from rest at a noncollinear configuration satisfies 
	\[
	\dot I_m(t)<0
	\qquad(0<t<T_*).
	\]
	At the initial instant $\dot I_m(0)=0$.
	If $T=T_*$ is a regular first syzygy, then
	$\dot I_m(T)<0$.
\end{theorem}

The second theorem deals with the type of the syzygy, i.e. which body is the middle body of the first syzygy. For general trajectories, this is hard to predict from the initial conditions. In the case of brake orbits with equal masses, however, this can be seen from the following theorem stating that the order of the side lengths is preserved until the time of the first syzygy:

\begin{theorem}\label{thm:ordering}
	For the $\alpha$-potential equations \eqref{eq:alpha_forces}, suppose $0<\alpha<2$, the masses are equal, and the initial triangle is scalene.
	There is a finite first regular syzygy time $T=T_*$, and the solution
	is collision-free on $[0,T]$.
	After one fixed relabeling, write
	\[
	a_1(0)>a_2(0)>a_3(0)>0.
	\]
	Then
	\begin{equation}\label{eq:orderstatement}
		a_1(t)>a_2(t)>a_3(t)>0\qquad(0\le t\le T).
	\end{equation}
	In particular, every initial strict ordering of the three pairwise
	distances is preserved.
\end{theorem}
\subsection{Related Research and Literature}
The occurrence of syzygies in the planar three-body problem has been studied extensively, particularly for solutions with zero angular momentum. \textsc{Montgomery} \cite{MontSyzygy} proved that every solution of the Newtonian three-body problem with zero angular momentum and negative energy encounters a syzygy, with the exception of the homothetic \textsc{Lagrange} solution, in which an equilateral triangle preserves its shape until triple collision.

Building on this result, \textsc{Moeckel}, \textsc{Montgomery} and \textsc{Venturelli} investigated the evolution of three bodies released from rest, referred to as \emph{brake orbits}, in their paper \cite{MMV}.

They introduced the \emph{brake-to-syzygy map}, which assigns to an initial configuration the configuration at its first syzygy. Using geometric and variational methods, they established continuity properties of this map and showed that its image contains a neighborhood of the binary collision configurations.

A central tool in their analysis is the \emph{shape sphere}, obtained by reducing translational and rotational symmetries and separating the overall size of the configuration from its shape. The resulting shape space is diffeomorphic to the two-sphere $S^2$, with collinear configurations corresponding to its equator. Consequently, syzygies can be interpreted as intersections of the reduced trajectories with this equator. Furthermore, the \textsc{Jacobi}--\textsc{Maupertuis} principle allows trajectories of fixed energy to be studied as geodesics of a suitable metric. By considering minimizing geodesics, the authors constructed brake orbits with prescribed first syzygies near binary collision configurations. For equal masses, they additionally proved that certain \textsc{Jacobi}--\textsc{Maupertuis} minimizers preserve the ordering of their mutual distances.

These results provide important insights into the geometry and dynamics of syzygies, although they primarily concern the classical Newtonian potential ($\alpha=1$). 

More recently, \textsc{Tsygvintsev} investigated the occurrence and timing of syzygies in the planar Newtonian three-body problem in \cite{Tsyg23,Tsyg24}. 
In addition to ordinary syzygies, where the positions of the three bodies become collinear, he considered \emph{velocity syzygies}, characterized by collinear velocity vectors. He proved that every global, collision-free solution with zero angular momentum and negative total energy exhibits infinitely many velocity syzygies. Furthermore, using comparison theorems for matrix \textsc{Riccati} differential equations, he derived explicit upper and lower bounds for the time of the first syzygy in terms of the initial conditions, masses, and bounds on the mutual distances. These results complement the qualitative existence theorems by providing quantitative estimates for when syzygies occur.

To prove our theorems, we combine results from \textsc{Montgomery}'s papers with the matrix method of \textsc{Tsygvintsev}, using mainly elementary matrix techniques. 

\subsection{Structure}
\label{sc:structure}
In Section~\ref{sc:preliminaries}, we define all necessary variables and introduce \textsc{Jacobi} coordinates along with our notation. Furthermore, we present the relevant matrices, their properties, and the relations between them. In Section~\ref{sc:theorem_inertia}, we prove Theorem~\ref{thm:inertia} using matrix dynamics and a \textsc{Riccati} equation.
This section also contains an elementary lemma reproving the finiteness of the time until the first syzygy. Section~\ref{sc:thm_ordering} is devoted to the proof of Theorem~\ref{thm:ordering}; we establish several auxiliary lemmata to prove our main result. Finally, in Section~\ref{sc:outlook}, we discuss applications of our results to other areas of mathematics and physics and outline ideas for future generalizations.

\section{Preliminaries}
\label{sc:preliminaries}
For the $\alpha$-potential model, we use the normalization $\alpha G=1$ whenever convenient. For the proof of Theorem \ref{thm:ordering} we take $m_1=m_2=m_3=1$. An arbitrary common mass $m>0$ and $G>0$ reduce
to these equations by $\widehat t=\sqrt{\alpha Gm}\,t$.
This positive time rescaling preserves ordering and derivative signs. No such normalization is used for the general central forces in Theorem~\ref{thm:inertia}.

\subsection{Conserved quantities}
Identifying the plane $\mathbb{R}^2$ with $\mathbb{C}$, the Euclidean inner
product and the two-dimensional determinant of any $z,w\in\mathbb{C}$ combine
into the single Hermitian product
\[
\bar z w = (z\cdot w) + \mathrm{i}\det(z,w),
\]
so that $z\cdot w=\operatorname{Re}(\bar z w)$ and
$\det(z,w)=\operatorname{Im}(\bar z w)$.

\begin{definition}[total angular momentum and kinetic energy]
	\label{def:ang_momentum_energy} \quad \\
	For a planar solution $q = (q_1, q_2, q_3)$, 
	the total angular momentum $L$ and the kinetic energy $\mathcal{T}$ are given by
	\[
	L = \sum_{i=1}^3 m_i \operatorname{Im}(\bar q_i \dot q_i)
	= \sum_{i=1}^3 m_i \det(q_i, \dot q_i),
	\qquad
	\mathcal{T} = \frac{1}{2} \sum_{i=1}^3 m_i |\dot q_i|^2.
	\]
\end{definition}
This definition leads to the classical conservation laws in the center of mass frame.
\begin{lemma}[Conservation laws]\label{lem:conservation}
	Let $q = (q_1, q_2, q_3) \in \mathbb{C}^3 \setminus \Delta$ be a solution of \eqref{eq:general_central_forces} starting from rest, i.e., $\dot q_i(0) = 0$ for $i \in \{1,2,3\}$. After translating the center of mass to the origin, the following identities hold throughout the interval of existence:
	\begin{align}
		\sum_{i=1}^3 m_i q_i(t) = 0 = \sum_{i=1}^3 m_i \dot q_i(t), \label{eq:com}\\
		L(t) = \sum_{i=1}^3 m_i \operatorname{Im}\bigl(\bar q_i(t)\dot q_i(t)\bigr) = 0. \label{eq:angmom}
    \end{align}
    For the $\alpha$-potential model \eqref{eq:alpha_forces}, we additionally have
    \begin{equation}
		E(t) = \mathcal{T}(t) - U\bigl(q(t)\bigr) = -U\bigl(q(0)\bigr) =: -U_0 < 0. \label{eq:energy}
	\end{equation}
\end{lemma}
\begin{proof}
	By symmetry and centrality of the pair forces, the total linear momentum $\sum_{i=1}^3 m_i \dot q_i$ and the total angular momentum $L$ are constants of motion. For the $\alpha$-potential model, translation and rotation invariance of $U$ also give the conservation of total energy $E = \mathcal{T} - U$. Evaluating these conserved quantities at $t = 0$, where $\dot q_i(0) = 0$ and $\mathcal{T}(0) = 0$, yields \eqref{eq:com}--\eqref{eq:energy} in the $\alpha$-potential case, and \eqref{eq:com}--\eqref{eq:angmom} in general.
\end{proof}
\subsection{Triangle geometry and moment of inertia}
\label{subsec:triangle_inertia}

The three bodies span a triangle, whose shape is described by its side lengths and its
oriented area. We define these quantities for every configuration, including collinear
ones and collisions, since they will also be evaluated at the first syzygy.

\begin{definition}[Triangle side lengths and area]
	\label{def:tri_length_area}
For cyclic indices \\ $(i, j, k) \in \{(1,2,3), (2,3,1), (3,1,2)\}$, we denote the side length opposite body~$i$ by
\begin{align}
	a_i := |q_k - q_j|, \label{def:a_i}
\end{align}

so that $a_1 = r_{23}$, $a_2 = r_{31}$, and $a_3 = r_{12}$. Twice the oriented area of the triangle is given by
\begin{align}
	B := \operatorname{Im}\bigl(\overline{(q_2 - q_1)}(q_3 - q_1)\bigr)
	= \det(q_2 - q_1, q_3 - q_1). \label{def:B_as_area}
\end{align}
\end{definition}

Since the initial configuration is noncollinear, a fixed reflection of the plane ensures $B(0) > 0$.

\begin{definition}[Maximal noncollinear interval ]\label{def:Tstar}
	For a classical solution $q = (q_1, q_2, q_3)$ of \eqref{eq:general_central_forces} with $B(0) > 0$, define
	\[
	T_* := \sup \bigl\{ t > 0 \,\big|\, q \text{ exists classically on } [0, t] \text{ with } q(s) \notin \Delta \text{ and } B(s) > 0 \text{ for all } s \in [0, t] \bigr\}.
	\]
	We call $[0, T_*)$ the \emph{maximal noncollinear initial interval}. If $T_* < \infty$ and $q$ extends smoothly to $T_*$ with $q(T_*) \notin \Delta$ and $B(T_*) = 0$, then $T_*$ is called the \emph{first regular syzygy time}.
\end{definition}
All calculations involving division by $B$ or invertible configuration matrices take place on the open-ended interval $[0, T_*)$.

\begin{definition}[Moment of inertia]\label{def:inertia}
	The \emph{moment of inertia} $I_m$ about the center of mass is defined by
	\begin{equation}\label{eq:inertia_defs}
		I_m = \sum_{i=1}^3 m_i |q_i|^2.
	\end{equation}
\end{definition}

\subsection{Jacobi coordinates}
\label{subsec:jacobi}

To decouple the center-of-mass constraint $\sum_{i=1}^3 m_i q_i = 0$, we identify $\mathbb{C} \cong \mathbb{R}^2$ with column vectors whenever matrix products are formed and introduce mass-weighted relative coordinates.

Let $M := m_1 + m_2 + m_3$ denote the \emph{total mass}. 

\begin{definition}[Mass-normalized Jacobi coordinates]\label{def:jacobi}
	Set $m_{12} := m_1 + m_2$, and define the reduced masses
	\[
	\mu := \frac{m_1 m_2}{m_{12}}, \qquad \nu := \frac{m_3 m_{12}}{M}.
	\]
	For a centered configuration $q = (q_1, q_2, q_3) \in \mathbb{C}^3$ with $\sum_{i=1}^3 m_i q_i = 0$, the \emph{Jacobi vectors} are given by
	\begin{equation}\label{eq:jacobi_vecs}
		r := q_2 - q_1, \qquad z := q_3 - \frac{m_1 q_1 + m_2 q_2}{m_{12}}.
	\end{equation}
	Assembling their mass-normalized counterparts into columns yields the $2 \times 2$ real \emph{configuration matrix}
	\begin{equation}\label{eq:Qdef}
		Q := \bigl(\sqrt{\mu}\,r \;\; \sqrt{\nu}\,z\bigr) \in \mathbb{R}^{2 \times 2}.
	\end{equation}
\end{definition}
Geometrically, $r$ is the relative vector of the pair $(1,2)$, and $z$ measures the position of body~$3$ relative to the pair's center of mass $\frac{m_1 q_1 + m_2 q_2}{m_{12}}$. The factors $\mu$ and $\nu$ are the corresponding reduced masses of the pair $(1,2)$ and of body~$3$ against the combined mass $m_{12}$, respectively.

\begin{lemma}\label{lem:jacobi_inverse}
	For any centered configuration $q = (q_1, q_2, q_3) \in \mathbb{C}^3$, the particle positions are recovered from $Q$ via
	\begin{equation}\label{eq:q_from_Q}
		q_i = Q d_i \qquad (i = 1, 2, 3),
	\end{equation}
	where the constant coefficient vectors $d_1, d_2, d_3 \in \mathbb{R}^2$ are given by
	\begin{equation}\label{eq:dvecs}
		d_1 = \begin{pmatrix} -\dfrac{m_2}{m_{12}\sqrt{\mu}} \\[6pt] -\dfrac{m_3}{M\sqrt{\nu}} \end{pmatrix}, \qquad
		d_2 = \begin{pmatrix} \dfrac{m_1}{m_{12}\sqrt{\mu}} \\[6pt] -\dfrac{m_3}{M\sqrt{\nu}} \end{pmatrix}, \qquad
		d_3 = \begin{pmatrix} 0 \\[6pt] \dfrac{m_{12}}{M\sqrt{\nu}} \end{pmatrix}.
	\end{equation}
	These vectors satisfy the mass-orthonormality relation
	\begin{equation}\label{eq:d_orthonormal}
		\sum_{i=1}^3 m_i d_i d_i^\top = \mathbf{1}_2,
	\end{equation}
	and consequently $\sum_{i=1}^3 m_i q_i d_i^\top = Q$.
\end{lemma}

\begin{proof}
	Let
	\[
	R_{12} := \frac{m_1 q_1 + m_2 q_2}{m_{12}}
	\]
	denote the center of mass of the pair $(1,2)$, so that $z = q_3 - R_{12}$. Using the center-of-mass condition $\sum_{i=1}^3 m_i q_i = 0$ and $M = m_{12} + m_3$, we obtain
	\[
	0 = m_{12} R_{12} + m_3 q_3 = m_{12} R_{12} + m_3 (R_{12} + z) = M R_{12} + m_3 z,
	\]
	which yields
	\[
	R_{12} = -\frac{m_3}{M} z, \qquad q_3 = R_{12} + z = \frac{m_{12}}{M} z.
	\]
	On the other hand, combining the definition of $R_{12}$ with $r = q_2 - q_1$ expresses $q_1$ and $q_2$ in terms of $R_{12}$ and $r$:
	\[
	q_1 = R_{12} - \frac{m_2}{m_{12}} r, \qquad q_2 = R_{12} + \frac{m_1}{m_{12}} r.
	\]
	Substituting $R_{12} = -(m_3/M)z$ and comparing with $Q = (\sqrt{\mu}\,r \;\; \sqrt{\nu}\,z)$ recovers $q_i = Q d_i$ for $i = 1, 2, 3$ with the coefficient vectors in \eqref{eq:dvecs}.
	
	To verify the mass-orthonormality relation \eqref{eq:d_orthonormal}, we compute the three independent entries of the symmetric $2 \times 2$ matrix $\sum_{i=1}^3 m_i d_i d_i^\top$ directly from \eqref{eq:dvecs}:
	\begin{align*}
		\sum_{i=1}^3 m_i (d_i)_1^2
		&= m_1 \frac{m_2^2}{m_{12}^2 \mu} + m_2 \frac{m_1^2}{m_{12}^2 \mu}
		= \frac{m_1 m_2 (m_1 + m_2)}{m_{12}^2 \mu}
		= \frac{m_1 m_2}{m_{12}\mu} = 1, \\[6pt]
		\sum_{i=1}^3 m_i (d_i)_2^2
		&= (m_1 + m_2) \frac{m_3^2}{M^2 \nu} + m_3 \frac{m_{12}^2}{M^2 \nu}
		= \frac{m_{12} m_3 (m_3 + m_{12})}{M^2 \nu}
		= \frac{m_{12} m_3}{M\nu} = 1, \\[6pt]
		\sum_{i=1}^3 m_i (d_i)_1 (d_i)_2
		&= -\frac{m_1 m_2 m_3}{m_{12} M \sqrt{\mu\nu}} + \frac{m_2 m_1 m_3}{m_{12} M \sqrt{\mu\nu}}
		= 0.
	\end{align*}
	Finally, substituting $q_i = Q d_i$ and applying \eqref{eq:d_orthonormal} gives
	\[
	\sum_{i=1}^3 m_i q_i d_i^\top
	= Q \left( \sum_{i=1}^3 m_i d_i d_i^\top \right)
	= Q \mathbf{1}_2 = Q. \qedhere
	\]
\end{proof}

Next, we can use the configuration matrix to calculate our system quantities:

\begin{lemma}\label{lem:Q_properties}
	For every centered configuration $q = (q_1, q_2, q_3) \in \mathbb{C}^3$, the configuration matrix $Q$ satisfies
	\begin{equation}\label{eq:inertia_tensor}
		\sum_{i=1}^3 m_i q_i q_i^\top = Q Q^\top, \qquad
		I_m = \operatorname{tr}(Q Q^\top) = \|Q\|_F^2,
	\end{equation} where $\|\cdot\|_F$ denotes the Frobenius norm,
	and
	\begin{equation}\label{eq:detQ}
		\det Q = \sqrt{\mu\nu}\, B,
	\end{equation}
	where $B$ is twice the area of the triangle, compare \eqref{def:B_as_area}.
	In particular, $Q(t)$ is invertible for all $t \in [0, T_*)$.
\end{lemma}

\begin{proof}
	Using \eqref{eq:q_from_Q} and \eqref{eq:d_orthonormal}, we compute
	\[
	\sum_{i=1}^3 m_i q_i q_i^\top = Q \left(\sum_{i=1}^3 m_i d_i d_i^\top\right) Q^\top = Q Q^\top.
	\]
Taking the trace gives
\[
I_m = \sum_{i=1}^3 m_i |q_i|^2 = \operatorname{tr}\!\left(\sum_{i=1}^3 m_i q_i q_i^\top\right) = \operatorname{tr}(Q Q^\top) = \|Q\|_F^2.
\] 
Finally, writing $z =q_3 -R_{12}= q_3 - q_1 - \frac{m_2}{m_{12}}(q_2 - q_1)$ and using the multilinearity and alternating property of the determinant map, we obtain
	\[
	\det Q = \sqrt{\mu\nu}\,\det(r, z) = \sqrt{\mu\nu}\,\det(q_2 - q_1, q_3 - q_1) = \sqrt{\mu\nu}\,B.
	\]
	Since $B(t) > 0$ on $[0, T_*)$, it follows that $\det Q(t) > 0$ and hence $Q(t)$ is invertible on $[0, T_*)$.
\end{proof}

The matrix identities in Lemma~\ref{lem:Q_properties} admit a direct geometric interpretation in terms of the mass distribution of the three bodies.

\begin{definition}[Configuration ellipse]\label{def:configuration_ellipse}
	For a centered configuration $q = (q_1, q_2, q_3) \in \mathbb{C}^3$ with configuration matrix $Q$, the \emph{configuration ellipse} is defined as the image of the unit circle under $Q$:
	\[
	\mathcal{E}(q) := \bigl\{ Q u \,\big|\, u \in \mathbb{R}^2,\, |u| = 1 \bigr\} \subset \mathbb{R}^2.
	\]
\end{definition}

By the singular value decomposition $Q = W \operatorname{diag}(\sigma_1, \sigma_2) V^\top$ with $W, V \in \mathrm{O}(2)$ and $\sigma_1 \ge \sigma_2 \ge 0$, the semi-axes of $\mathcal{E}(q)$ are given by the singular values $\sigma_1, \sigma_2$ (whose squares are the eigenvalues of $Q Q^\top$). In particular:
\begin{itemize}
	\item The sum of the squared semi-axes equals the moment of inertia:
	\[
	\sigma_1^2 + \sigma_2^2 = \operatorname{tr}(Q Q^\top) = I_m.
	\]
	\item The enclosed area of $\mathcal{E}(q)$ is proportional to the triangle area:
	\[
	\operatorname{Area}\bigl(\mathcal{E}(q)\bigr) = \pi |\det Q| = \pi \sqrt{\mu\nu}\, |B|.
	\]
\end{itemize}

Thus, on $[0,T_*)$, $\mathcal{E}\bigl(q(t)\bigr)$
is a nondegenerate ellipse. In fact, we will show
in Corollary~\ref{cor:spatial_moment} that the
derivative of the second mass moment tensor
$QQ^\top$ is strictly negative definite.
Consequently, both squared semi-axes
$\sigma_1^2$ and $\sigma_2^2$ decrease strictly,
not merely their sum $I_m$.\\

A regular syzygy ($B(T_*) = 0$ with $q(T_*) \notin \Delta$) corresponds precisely to the degeneration of the configuration ellipse into a line segment of positive length $\sigma_1 > 0$, $\sigma_2 = 0$, whereas a triple collision corresponds to $\sigma_1 = \sigma_2 = 0$.

\begin{remark}\label{rem:mass_orthonormality}
	On the center-of-mass subspace $\mathcal{X} = \{q \in \mathbb{C}^3 \mid \sum_{i=1}^3 m_i q_i = 0\}$, the natural Riemannian structure is given by the \emph{mass metric} (or kinetic-energy inner product)
	\[
	\langle q, \tilde q \rangle_m := \sum_{i=1}^3 m_i \operatorname{Re}(\bar q_i \tilde q_i) = \sum_{i=1}^3 m_i\, q_i^\top \tilde q_i,
	\]
	under which the moment of inertia and the kinetic energy take the metric forms $I_m(q) = \langle q, q \rangle_m$ and $\mathcal{T} = \frac{1}{2}\langle \dot q, \dot q \rangle_m$. On $\mathbb{R}^3$, the corresponding weighted inner product $(u, v)_m := \sum_{i=1}^3 m_i u_i v_i$ turns the two rows of $D = (d_1 \;\; d_2 \;\; d_3)$ into an orthonormal basis of the two-dimensional subspace $\{u \in \mathbb{R}^3 \mid \sum_{i=1}^3 m_i u_i = 0\}$, which is the precise algebraic content of \eqref{eq:d_orthonormal}.

	We equip $\mathbb{R}^{2 \times 2}$ with the Euclidean Frobenius inner product $\langle Q, \tilde Q \rangle_F := \operatorname{tr}(Q \tilde Q^\top)$.
	Consider the coordinate map $Q \mapsto Qd := (Qd_1, Qd_2, Qd_3)$ between $(\mathbb{R}^{2 \times 2}, \langle \cdot, \cdot \rangle_F)$ and $(\mathcal{X}, \langle \cdot, \cdot \rangle_m)$. With \eqref{eq:d_orthonormal} we obtain
	\begin{align*}
		\langle Qd, \tilde Q d \rangle_m
		&= \sum_{i=1}^3 m_i (Q d_i)^\top (\tilde Q d_i)
		= \sum_{i=1}^3 m_i \operatorname{tr}\!\bigl((Q d_i)(\tilde Q d_i)^\top\bigr) \\
		&= \operatorname{tr}\!\left(Q \left(\sum_{i=1}^3 m_i d_i d_i^\top\right) \tilde Q^\top\right)
		= \operatorname{tr}(Q \tilde Q^\top)
		= \langle Q, \tilde Q \rangle_F.
	\end{align*}
	Thus, the coordinate map is a linear isometry. In particular, setting $\tilde q = q$ and $\tilde q = \dot q$ recovers $I_m(q) = \|Q\|_F^2$ and $\mathcal{T} = \frac{1}{2}\|\dot Q\|_F^2$. 
	
	Geometrically, rescaling by $\sqrt{\mu}$ and $\sqrt{\nu}$ flattens the mass-weighted metric into the standard Euclidean metric: it maps the ellipsoid of configurations with unit moment of inertia $\{q \in \mathcal{X} \mid I_m(q) = 1\}$ onto the Euclidean unit sphere $\{Q \in \mathbb{R}^{2 \times 2} \mid \|Q\|_F = 1\}$, and eliminates all mass weights from the spatial inertia tensor $\sum_{i=1}^3 m_i q_i q_i^\top = Q Q^\top$.
\end{remark}
In summary, although the planar configuration $q = (q_1, q_2, q_3) \in \mathbb{C}^3$ involves six real coordinates, the center-of-mass constraint imposes two real linear relations, leaving four genuine internal degrees of freedom. We will work in this space and do not apply any other reductions, e.g. we do not use the rotational symmetry of the potential.

\subsection{Matrix formulation}
To express the equations of motion in terms of the configuration matrix $Q$, we first introduce the relative coefficient vectors and the matrix $\mathcal{S}$ associated with attractive central pair forces. \\
As in Remark~\ref{rem:mass_orthonormality} we define $$\mathcal{X} := \{q \in \mathbb{C}^3 \mid \sum_{i=1}^3 m_i q_i = 0\}$$

\begin{definition}\label{def:S_matrix}
	For $1 \le i < j \le 3$, define the relative coefficient vectors
	\begin{equation}\label{eq:cij}
		c_{ij} := d_j - d_i \in \mathbb{R}^2.
	\end{equation}
	For a collision-free configuration $q \in \mathcal{X} \setminus \Delta$ with mutual distances $r_{ij}> 0$, define $\mathcal{S}\in \operatorname{Mat}(2,\mathbb{R})$
	\begin{equation}\label{eq:matrix_S}
		\mathcal{S} := \sum_{1 \le i < j \le 3} \kappa_{ij}(r_{ij})\, c_{ij} c_{ij}^\top.
	\end{equation}
\end{definition}

Using the matrix $\mathcal{S}$, we can write the equations of motion for the matrix in a compact way:

\begin{lemma}\label{lem:newton_matrix}
Every collision-free solution $q(t)$ of \eqref{eq:general_central_forces} in the planar three-body problem satisfies
	\begin{equation}\label{eq:newton_Q}
		\ddot{Q} = -Q\mathcal{S}.
	\end{equation}
\end{lemma}

\begin{proof}
	First, note that by $q_i = Q d_i$ and \eqref{eq:cij}, the relative position vectors can be written in matrix form as
	\[
	q_j - q_i = Q d_j - Q d_i = Q(d_j - d_i) = Q c_{ij}.
	\]
	Newton's equations of motion for the three bodies read
	\[
	m_i \ddot{q}_i = \sum_{j \neq i} \kappa_{ij}(r_{ij}) (q_j - q_i), \qquad i = 1, 2, 3.
	\]
	Differentiating $Q = \sum_{i=1}^3 m_i q_i d_i^\top$ twice with respect to time and substituting Newton's equations yields
	\[
	\ddot{Q}
	= \sum_{i=1}^3 m_i \ddot{q}_i d_i^\top
	= \sum_{i=1}^3 \sum_{j \neq i} \kappa_{ij}(r_{ij}) (q_j - q_i) d_i^\top.
	\]
	This double sum consists of six terms, which naturally group into the three unordered pairs $1 \le i < j \le 3$. For a single pair $(i,j)$, the two corresponding terms combine to
	\begin{align*}
		\kappa_{ij}(r_{ij}) \bigl[(q_j - q_i) d_i^\top + (q_i - q_j) d_j^\top\bigr]
        &= \kappa_{ij}(r_{ij}) (q_j - q_i)(d_i - d_j)^\top \\[6pt]
        &= -\kappa_{ij}(r_{ij}) (q_j - q_i)(d_j - d_i)^\top \\[6pt]
        &= -\kappa_{ij}(r_{ij}) Q c_{ij} c_{ij}^\top.
	\end{align*}
	Summing this identity over all three pairs $1 \le i < j \le 3$ gives
	\[
	\ddot{Q}
	= -\sum_{1 \le i < j \le 3} \kappa_{ij}(r_{ij}) Q c_{ij} c_{ij}^\top
    = -Q \left( \sum_{1 \le i < j \le 3} \kappa_{ij}(r_{ij}) c_{ij} c_{ij}^\top \right)
	= -Q\mathcal{S}. \qedhere
	\]
\end{proof}

\begin{lemma}\label{lem:S_pos_def}
	For every  $q \in \mathcal{X} \setminus \Delta$, the matrix $\mathcal{S}$ is symmetric and strictly positive definite.
\end{lemma}

\begin{proof}
	$\mathcal{S}$ is symmetric directly from the definition.
	
	To prove strict positive definiteness, let $w \in \mathbb{R}^2$ be an arbitrary vector and evaluate the quadratic form $w^\top \mathcal{S} w$:
	\[
	w^\top \mathcal{S} w
	= \sum_{1 \le i < j \le 3} \kappa_{ij}(r_{ij}) w^\top c_{ij} c_{ij}^\top w
    = \sum_{1 \le i < j \le 3} \kappa_{ij}(r_{ij}) (c_{ij}^\top w)^2.
    \]
    Since $\kappa_{ij}(r_{ij}) > 0$ for all $i < j$, every summand is non-negative, whence $w^\top \mathcal{S} w \ge 0$. Moreover, $w^\top \mathcal{S} w = 0$ holds if and only if each term vanishes individually, which requires
	\[
	c_{12}^\top w = 0, \qquad c_{13}^\top w = 0, \qquad c_{23}^\top w = 0.
	\]
	From the explicit formulas \eqref{eq:dvecs} for $d_1, d_2, d_3$, we compute the first two relative coefficient vectors:
	\[
	c_{12} = d_2 - d_1 = \begin{pmatrix} 1/\sqrt{\mu} \\[4pt] 0 \end{pmatrix}, \qquad
	c_{13} = d_3 - d_1 = \begin{pmatrix} m_2/(m_{12}\sqrt{\mu}) \\[4pt] 1/\sqrt{\nu} \end{pmatrix}.
	\]
	Hence $c_{12}$ and $c_{13}$ are linearly independent and span $\mathbb{R}^2$. Consequently, the conditions $c_{12}^\top w = 0$ and $c_{13}^\top w = 0$ imply $w = 0$. Therefore, $w^\top \mathcal{S} w > 0$ for all $w \neq 0$, showing that $\mathcal{S}$ is strictly positive definite.
\end{proof}

\section{Proof of Theorem \ref{thm:inertia}}
\label{sc:theorem_inertia}

We now prove Theorem~\ref{thm:inertia} using the matrix
formulation introduced in Section~\ref{sc:preliminaries}.
Our argument is based on the matrix Riccati equation
associated with the configuration matrix $Q$, following
the matrix approach of \textsc{Tsygvintsev}
\cite{Tsyg24}.

Recall from Lemma~\ref{lem:newton_matrix} that
\begin{equation}
	\ddot Q=-Q\mathcal S,
\end{equation}
where $\mathcal S$ is symmetric and strictly positive
definite by Lemma~\ref{lem:S_pos_def}.

By Definition~\ref{def:Tstar} and
Lemma~\ref{lem:Q_properties}, the matrix $Q(t)$ is
invertible on $[0,T_*)$. Furthermore, the initial
rest condition implies $\dot Q(0)=0$.

\subsection{Matrix Riccati equation}

We first introduce an auxiliary matrix and establish
its symmetry.

\begin{definition}\label{def:C_matrix}
	For $t\in[0,T_*)$, define
	\begin{equation}\label{eq:C_definition}
		C(t):=Q(t)^{-1}\dot Q(t).
	\end{equation}
	Equivalently, $\dot Q=QC$, i.e. $C$ describes the coordinates of the vectors $\sqrt{\mu} \dot{r}, \sqrt{\nu} \dot{z}$ in the basis $\sqrt{\mu} r, \sqrt{\nu} z$.
\end{definition}

\begin{lemma}\label{lem:C_symmetric}
	The matrix $C(t)$ is symmetric for every
	$t\in[0,T_*)$.
\end{lemma}

\begin{proof}
	Consider
	\begin{equation}
		W:=\dot Q Q^\top-Q\dot Q^\top.
	\end{equation}
	Differentiating and using
	\eqref{eq:newton_Q}, we obtain
	\begin{align}
		\dot W
		&=\ddot Q Q^\top-Q\ddot Q^\top
		=-Q\mathcal S Q^\top
		+Q\mathcal S^\top Q^\top
		=0,
	\end{align}
	since $\mathcal S=\mathcal S^\top$.
	
	Hence $W$ is constant. From $\dot Q(0)=0$
	we have $W(0)=0$, and therefore $W(t)=0$
	throughout $[0,T_*)$.
	
	Multiplying by $Q^{-1}$ from the left
	and by $Q^{-\top}$ from the right gives
	\begin{align}
		0
		&=Q^{-1}WQ^{-\top}=Q^{-1}\dot Q-\dot Q^\top Q^{-\top}
		=C-C^\top.
	\end{align}
	Thus $C=C^\top$.
\end{proof}

\begin{remark}\label{rem:C_interpretation}
	The matrices $\dot Q$ and $C$ carry the same information in two forms. The columns of
	$\dot Q$ are the mass-normalized Jacobi velocities $\sqrt{\mu}\,\dot r$ and
	$\sqrt{\nu}\,\dot z$, and $\dot q_i = \dot Q d_i$ recovers all three velocities.
	The matrix $C$ expresses these velocities in the basis formed by the current
	configuration. Equivalently, setting $A := \dot Q Q^{-1} = QCQ^{-1}$, the velocities are the values of
	the linear vector field $x \mapsto Ax$ at the positions of the bodies,
	\[
	\dot q_i = A\, q_i \qquad (i = 1, 2, 3).
	\]
 Since $A$ and $C$
	are similar, $\operatorname{tr} C = \operatorname{tr} A = \dot B/B$ is the relative rate
	of change of the area, cf.~\eqref{eq:B_log_derivative}.
	
	The symmetry of $C$ expresses the vanishing of the angular momentum: arguing as in
	Remark~\ref{rem:mass_orthonormality}, $Q\dot Q^\top = \sum_{i=1}^3 m_i\, q_i \dot q_i^\top$,
	hence
	\[
	W = \dot Q Q^\top - Q \dot Q^\top
	= \sum_{i=1}^3 m_i \bigl(\dot q_i q_i^\top - q_i \dot q_i^\top\bigr)
	= L \begin{pmatrix} 0 & -1 \\ 1 & 0 \end{pmatrix}.
	\]
	Thus $\dot W = 0$ is the conservation of angular momentum, and, since $Q$ is invertible,
	$C$ is symmetric if and only if $L = 0$.
\end{remark}

The symmetry of $C$ allows us to establish a
stronger property.

\begin{lemma}\label{lem:C_negative}
	For every $t\in(0,T_*)$, the matrix $C(t)$
	is strictly negative definite.
\end{lemma}

\begin{proof}
	Differentiating $Q^{-1}Q=\Id$ yields
	\begin{equation}
		\frac{\dd}{\dd t}Q^{-1}
		=-Q^{-1}\dot Q Q^{-1}.
	\end{equation}
	
	Hence, differentiating \eqref{eq:C_definition}
	and applying \eqref{eq:newton_Q}, we obtain
	\begin{align*}
		\dot C
		&=-Q^{-1}\dot Q Q^{-1}\dot Q
		+Q^{-1}\ddot Q
		=-C^2-\mathcal S.
	\end{align*}
	
	Thus $C$ satisfies the matrix \textsc{Riccati} equation
	\begin{equation}\label{eq:C_riccati}
		\dot C=-C^2-\mathcal S,
			\qquad C(0)=0.
	\end{equation}
	
	Let $w\in\R^2\setminus\{0\}$ be arbitrary.
	Integrating \eqref{eq:C_riccati}
	from $0$ to $t\in(0,T_*)$ gives
	\begin{equation*}
		w^\top C(t)w
		=-\int_0^t
		\left(
		w^\top C(s)^2w+
		w^\top\mathcal S(s)w
		\right)\dd s.
	\end{equation*}
	
	By Lemma~\ref{lem:C_symmetric},
	\begin{equation*}
		w^\top C(s)^2w=|C(s)w|^2\geq0.
	\end{equation*}
	Moreover, Lemma~\ref{lem:S_pos_def} implies
	\begin{equation*}
		w^\top\mathcal S(s)w>0.
	\end{equation*}
	
	Putting these results together, we obtain
	\begin{equation}\label{eq:C_negative}
			w^\top C(t)w
			=-\int_0^t
			\left(
			|C(s)w|^2+
			w^\top\mathcal S(s)w
			\right)\dd s<0.
	\end{equation}
	
	Since $w\neq0$ was arbitrary, $C(t)$ is
	strictly negative definite.
\end{proof}

\begin{corollary}[Spatial moment contraction]
    \label{cor:spatial_moment}
    The second mass moment tensor
    \[
        \mathcal D(t)
        :=\sum_{i=1}^3 m_i q_i(t)q_i(t)^\top
        =Q(t)Q(t)^\top
    \]
    has a strictly negative definite time derivative
    for every $0<t<T_*$.
\end{corollary}

\begin{proof}
    Differentiating and using $\dot Q=QC$
    together with $C=C^\top$, we obtain
    \[
        \dot{\mathcal D}
        =\dot Q Q^\top+Q\dot Q^\top
        =2QCQ^\top.
    \]
    Since $C$ is strictly negative definite by
    Lemma~\ref{lem:C_negative} and $Q$ is invertible
    on $[0,T_*)$, the matrix
    $\dot{\mathcal D}(t)$ is strictly negative
    definite for every $0<t<T_*$.
\end{proof}

\begin{corollary}[Strict contraction of the configuration ellipse]
    \label{cor:ellipse_axes}
    Let $\sigma_1(t)\geq\sigma_2(t)>0$ denote the
    semi-axes of the configuration ellipse.
    Then both $\sigma_1$ and $\sigma_2$ are strictly
    decreasing on $[0,T_*)$.
\end{corollary}

\begin{proof}
    By Corollary~\ref{cor:spatial_moment},
$\dot{\mathcal D}(t)$ is strictly negative definite
for every $t\in(0,T_*)$.

Hence, for any $0\leq s<t<T_*$,
\[
    \mathcal D(s)-\mathcal D(t)
    =-\int_s^t\dot{\mathcal D}(u)\,\mathrm du
\]
is strictly positive definite.

Consequently, there exists $\varepsilon>0$
such that
\[
    \mathcal D(s)-\mathcal D(t)-\varepsilon\Id
\]
is positive semidefinite.

    By the min-max characterization of the
    eigenvalues of symmetric matrices,
    \[
        \lambda_j(\mathcal D(s))
        \geq\lambda_j(\mathcal D(t))
        +\varepsilon,
        \qquad j=1,2.
    \]
    Since $\lambda_j(\mathcal D)=\sigma_j^2$
    and both singular values are positive,
    we obtain
    \[
        \sigma_j(s)>\sigma_j(t),
        \qquad j=1,2.
    \]
\end{proof}

We now use the negative definiteness of $C$ and $\dot{\mathcal{D}}$ to establish the strict monotonicity of $I_m$.

\begin{lemma}\label{lem:inertia_decrease}
	The moment of inertia satisfies
	\begin{equation}
		\dot I_m(t)<0
		\qquad (0<t<T_*).
	\end{equation}
\end{lemma}

\begin{proof}
	By Lemma~\ref{lem:Q_properties},
	\begin{equation}
		I_m=\tr(QQ^\top) = \tr(\mathcal{D}), \text{ so } \dot{I}_m = \tr(\dot{\mathcal{D}}) \label{eq:Im_derivative}
	\end{equation}
	and the statement follows from Corollary~\ref{cor:spatial_moment}.
\end{proof}

\subsection{Finiteness of the first endpoint}

It remains to show that $T_*<\infty$.
For this purpose, we use the oriented
double area $B$ introduced in
Definition~\ref{def:tri_length_area}. For $\alpha = 1$, this result was proved by Montgomery in~\cite{MontSyzygy}.

\begin{lemma}\label{lem:sqrtB_concave}
	The function $\sqrt{B(t)}$ is strictly
	decreasing and strictly concave on
	$(0,T_*)$. In particular, $T_*<\infty$.
\end{lemma}

\begin{proof}
	By Lemma~\ref{lem:Q_properties}, $\det Q=\sqrt{\mu\nu}\,B$, where $\mu,\nu>0$ are constant.
	
	Since $B(t)>0$ on $[0,T_*)$,
	the matrix $Q(t)$ is invertible and
	\textsc{Jacobi}'s formula for the derivative of a determinant gives
	\begin{align}
		\frac{\dd}{\dd t}\det Q
		&=(\det Q)\tr(Q^{-1}\dot Q)
		=(\det Q)\tr C,
	\text{ which is equivalent to } \label{eq:B_log_derivative}
		\frac{\dot B}{B}=\tr C
	\end{align}

	By Lemma~\ref{lem:C_negative},
	both eigenvalues of $C(t)$ are strictly
	negative for $t>0$. Thus
	\begin{equation*}
		\dot B(t)=B(t)\tr C(t)<0.
	\end{equation*}
	
	In particular,
	\begin{equation*}
		\frac{\dd}{\dd t}\sqrt B
		=\frac12\sqrt B\,\tr C<0.
	\end{equation*}
	
	Differentiating once more, we obtain
	\begin{equation*}
		\frac{1}{\sqrt{B}} \frac{d^2}{dt^2}\sqrt{B}
		= \frac{1}{2}\tr\dot C + \frac{1}{4}(\tr C)^2.
	\end{equation*}
	
	Substituting the Riccati equation \eqref{eq:C_riccati} yields
	\begin{equation}
		\frac{1}{\sqrt{B}} \frac{d^2}{dt^2}\sqrt{B}
		= -\frac{1}{2}\tr\mathcal S - \frac{1}{2}\tr(C^2) + \frac{1}{4}(\tr C)^2.
	\end{equation}
	
	Let $\lambda_1,\lambda_2 \in \R$ denote the eigenvalues of the symmetric matrix $C$. Then
	\begin{equation*}
		\tr C = \lambda_1 + \lambda_2,
		\qquad
		\tr(C^2) = \lambda_1^2 + \lambda_2^2.
	\end{equation*}
	
	Hence
	\begin{align}
		\label{eq:sqrtB_concavity}
		\frac{1}{\sqrt{B}} \frac{d^2}{dt^2}\sqrt{B}
		&= -\frac{1}{2}\tr\mathcal S - \frac{1}{2}(\lambda_1^2 + \lambda_2^2) + \frac{1}{4}(\lambda_1 + \lambda_2)^2 \\
		&= -\frac{1}{2}\tr\mathcal S - \frac{1}{4}(\lambda_1 - \lambda_2)^2 
		< 0,
	\end{align}
	where the strict inequality follows from the positive definiteness of $\mathcal S$.
Since $\sqrt{B(t)} > 0$ on $(0,T_*)$, multiplying \eqref{eq:sqrtB_concavity} by $\sqrt{B}$ yields $\frac{d^2}{dt^2}\sqrt{B(t)} < 0$ for all $t \in (0,T_*)$. 

The function $t \mapsto \sqrt{B(t)}$ is therefore strictly concave on $(0,T_*)$.
	
	To prove that $T_*$ is finite,
	fix any $s\in(0,T_*)$.
	Since a concave function lies below its tangent lines and $\frac{d}{ds}\sqrt{B(s)} = \frac{\dot B(s)}{2\sqrt{B(s)}}$, we obtain
	\begin{equation}
		\sqrt{B(t)}
		\leq \sqrt{B(s)} + \frac{\dot B(s)}{2\sqrt{B(s)}}(t-s)
		\qquad (s \leq t < T_*).
	\end{equation}

	Since $\dot B(s)<0$, the right-hand
	side becomes nonpositive at the
	finite time
	\begin{equation}
		t_0=s+\frac{2B(s)}{-\dot B(s)}.
	\end{equation}
	
	However, $B(t)>0$ on $[0,T_*)$.
	Thus the interval cannot extend
	beyond $t_0$, giving
	\begin{equation}
			T_*\leq s+\frac{2B(s)}{-\dot B(s)}
			<\infty.
		\label{eq:Tstar_bound}
	\end{equation}
\end{proof}

\subsection{Proof of Theorem~\ref{thm:inertia}}
Combining the auxiliary results established above, we are now in a position to prove the first theorem.

\begin{proof}[Proof of Theorem~\ref{thm:inertia}]
	At the initial instant, the brake condition $\dot Q(0) = 0$ implies \linebreak $\dot I_m(0) = 2\tr(\dot Q(0)Q(0)^\top) = 0$. On the open interval $(0, T_*)$, Lemma~\ref{lem:inertia_decrease} yields $\dot I_m(t) < 0$, while Lemma~\ref{lem:sqrtB_concave} guarantees that $T_* < \infty$. It therefore remains only to verify that $\dot I_m(T_*) < 0$ whenever $T_*$ is a regular syzygy.
	
	To this end, suppose that $T_*$ is a regular syzygy and fix an arbitrary time $s \in (0, T_*)$. Since $C(s)$ is symmetric and strictly negative definite with eigenvalues $\lambda_1 \le \lambda_2 < 0$, setting $c_s := -\lambda_2 > 0$ yields $C(s) \le -c_s\Id$ in the ordering of symmetric matrices, i.e., $v^\top C(s) v \le -c_s |v|^2$ for all $v \in \mathbb{R}^2$. Moreover, the Riccati equation~\eqref{eq:C_riccati} gives
	\[
	\dot C(t) = -C(t)^2 - \mathcal S(t) < 0 \qquad (0 < t < T_*),
	\]
	so $t \mapsto C(t)$ is monotonically decreasing on $(0, T_*)$. In particular,
	\[
	C(t) \le C(s) \le -c_s\Id \qquad (s \le t < T_*).
	\]
	Substituting this bound into~\eqref{eq:Im_derivative} and recalling that $I_m(t) = \tr(Q(t)Q(t)^\top)$, we obtain
	\begin{align*}
		\dot I_m(t)
		&= 2\tr\bigl(Q(t)C(t)Q(t)^\top\bigr) 
		\le -2c_s\tr\bigl(Q(t)Q(t)^\top\bigr)
		= -2c_s I_m(t) \qquad (s \le t < T_*).
	\end{align*}
	
	Because $T_*$ is a regular syzygy, the three bodies remain pairwise distinct at $t = T_*$, so the solution extends smoothly through $T_*$ with $I_m(T_*) > 0$. Passing to the limit $t \nearrow T_*$ in the inequality above yields
	\[
	\dot I_m(T_*) \le -2c_s I_m(T_*) < 0. \qedhere
	\]
\end{proof}

\section{Proof of Theorem \ref{thm:ordering}}
\label{sc:thm_ordering}

In the equal-mass setting $m_1 = m_2 = m_3 = 1$, the center-of-mass constraint reads $q_1 + q_2 + q_3 = 0$. As above let $\alpha G=1$.

Assume as in the theorem, that we have initial side lengths satisfying $a_1(0)>a_2(0)>a_3(0)>0$. We give a short sketch of the proof, all these quantities will be defined formally below. \\ The proof proceeds in three steps. First, we show that on $[0,T_*)$ the sign of
each squared-side difference $f_i$ is preserved (Section~\ref{subsec:ordering_open});
this holds for all $\alpha>0$ and rests on two identities: the torque equation for
the internal angular momentum $K_i$ of pair $j,k$  and a kinematic relation between $f_i$, $B$ and $K_i$.
Second, for $0<\alpha<2$ we bound all side lengths away from zero, which shows that
$T_*$ is a regular syzygy (Sections~\ref{subsec:distance_lower_bound}
and~\ref{subsec:continuation_syzygy}). Third, we show that the ordering does not
degenerate at $T_*$ itself (Section~\ref{subsec:ordering_at_syzygy}).

Sections~\ref{subsec:tri_geo}--\ref{subsec:kinematic_relation} provide the tools for the first step. For each body~$i$ we introduce local coordinates adapted to the opposite side: the side vector $r_i$ and the median $s_i$. In these coordinates the double area $B$, the squared-side difference $f_i$ and the moment of inertia $I_m$ take a simple form, and the rotation of the side is measured by a single scalar $K_i$ (Section~\ref{subsec:tri_geo}). Section~\ref{subsec:rot_dynamics} shows that the sign of $f_i$ determines how $K_i$ evolves, and Section~\ref{subsec:kinematic_relation} shows conversely that $K_i$ drives the evolution of $f_i$ relative to $B$.

\subsection{Triangle geometry in side-median coordinates}\label{subsec:tri_geo}

Comparing the two sides $a_j$ and $a_k$ adjacent to body~$i$ amounts to asking which endpoint of the opposite side lies closer to body~$i$. It is therefore natural to describe the configuration relative to that side. We associate such a coordinate frame to each cyclic ordering of the three bodies.

\begin{definition}\label{def:side_frame}
	For any cyclic permutation $(i, j, k)$ of $(1, 2, 3)$, define the \emph{side vector} $r_i$ and the \emph{median vector} $s_i$ by
	\begin{equation}\label{eq:side_frame_def}
		r_i := q_k - q_j, \qquad s_i := q_i - \frac{q_j + q_k}{2}.
	\end{equation}
	
	Moreover, define the \emph{squared-side difference} $f_i$ and the \emph{internal angular momentum of the pair $j,k$}, cf. Remark \ref{rem:Ki_physical}, $K_i$ by

	\begin{equation}\label{eq:fi_Ki_def}
		f_i := a_j^2 - a_k^2, \qquad K_i := \operatorname{Im}(\bar{r}_i \dot{r}_i).
	\end{equation}
\end{definition}

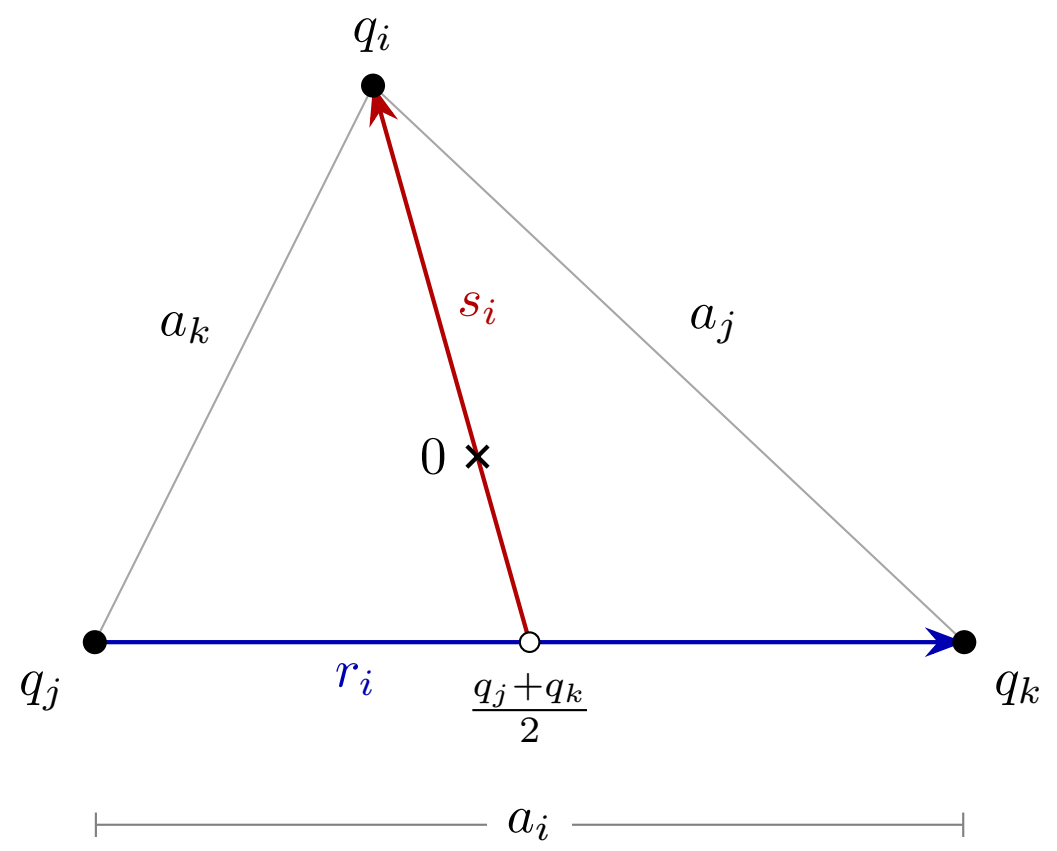
\begin{figure}[ht]
	\centering
	\begin{tikzpicture}[scale=1.15,
		vec/.style={-{Stealth[length=2.6mm]}, thick},
		body/.style={circle, fill, inner sep=1.6pt},
		lbl/.style={font=\small}]
		\coordinate (qj) at (0,0);
		\coordinate (qk) at (5,0);
		\coordinate (qi) at (1.6,3.2);
		\coordinate (M)  at ($(qj)!0.5!(qk)$);
		\coordinate (O)  at ($(M)!1/3!(qi)$);
		\draw[gray!70] (qi) -- node[lbl, above left, black] {$a_k$} (qj);
		\draw[gray!70] (qi) -- node[lbl, above right, black] {$a_j$} (qk);
		\draw[vec, blue!70!black] (qj) -- node[lbl, below, pos=0.3] {$r_i$} (qk);
		\draw[vec, red!70!black] (M) -- node[lbl, right, pos=0.6] {$s_i$} (qi);
		\node[body, label={[lbl]above:$q_i$}] at (qi) {};
		\node[body, label={[lbl]below left:$q_j$}] at (qj) {};
		\node[body, label={[lbl]below right:$q_k$}] at (qk) {};
		\node[circle, draw, fill=white, inner sep=1.3pt,
		label={[lbl]below:$\tfrac{q_j+q_k}{2}$}] at (M) {};
		\node[cross out, draw, inner sep=1.6pt, thick,
		label={[lbl]left:$0$}] at (O) {};
		\draw[|-|, gray] ($(qj)+(0,-1.05)$) -- node[lbl, fill=white, text=black] {$a_i$} ($(qk)+(0,-1.05)$);
	\end{tikzpicture}
	\caption{Side-median coordinates for a cyclic triple $(i,j,k)$. The side vector
		$r_i = q_k - q_j$ runs along the side of length $a_i$ opposite body~$i$, and the
		median vector $s_i$ joins the midpoint of that side to body~$i$. The center of mass
		$0$ lies on the median, with $q_i = \tfrac23 s_i$. In the configuration shown,
		$a_j > a_k$, i.e.\ $f_i > 0$.}
	\label{fig:side_frame}
\end{figure}

Geometrically, $r_i$ runs along the side opposite body~$i$, whereas $s_i$ joins the midpoint of that side to body~$i$.

\begin{lemma}
	\label{lem:side_frame_inverse}
	For every cyclic permutation $(i, j, k)$ of $(1, 2, 3)$ and every configuration satisfying $q_i + q_j + q_k = 0$, the particle positions are recovered from $(r_i, s_i)$ by
	\begin{equation}\label{eq:side_frame_inverse}
		q_i = \frac{2}{3} s_i, \qquad
		q_j = -\frac{1}{3} s_i - \frac{1}{2} r_i, \qquad
		q_k = -\frac{1}{3} s_i + \frac{1}{2} r_i.
	\end{equation}
\end{lemma}

\begin{proof}
	Using the center-of-mass condition $q_i + q_j + q_k = 0$, we have $q_j + q_k = -q_i$. Substituting this into the definition of $s_i$ in~\eqref{eq:side_frame_def} gives
	\[
	s_i = q_i - \frac{-q_i}{2} = \frac{3}{2} q_i,
	\]
	which yields $q_i = \frac{2}{3} s_i$ and consequently
	\[
	q_j + q_k = -\frac{2}{3} s_i.
	\]
	Combining this relation with $q_k - q_j = r_i$ and solving the resulting linear system for $q_j$ and $q_k$ gives~\eqref{eq:side_frame_inverse}.
\end{proof}

We next express twice the oriented triangle area and the moment of inertia in terms of the new coordinates $(r_i, s_i)$.

\begin{proposition}[Geometric invariants and moment of inertia]\label{prop:side_frame_geometry}
	For any cyclic permutation $(i, j, k)$ of $(1, 2, 3)$, the following identities hold:
	\begin{enumerate}
		\item The oriented double area $B$ and the squared-side difference $f_i$ are given by
		\begin{equation}\label{eq:B_fi_re_im}
			B = \operatorname{Im}(\bar{r}_i s_i), \qquad f_i = -2\operatorname{Re}(\bar{r}_i s_i),
		\end{equation}
		or equivalently, in combined complex form,
		\begin{equation}\label{eq:ri_si_product}
			\bar{r}_i s_i = -\frac{1}{2} f_i + \mathrm{i} B.
		\end{equation}
		\item The moment of inertia $I_m$ decouples into
		\begin{equation}\label{eq:Im_side_frame}
			I_m = \frac{2}{3} |s_i|^2 + \frac{1}{2} |r_i|^2.
		\end{equation}
	\end{enumerate}
\end{proposition}

\begin{proof}
	By Lemma~\ref{lem:side_frame_inverse}, the relative position vectors from body~$i$ to bodies~$j$ and~$k$ read
	\[
	q_j - q_i = -s_i - \frac{1}{2} r_i, \qquad
	q_k - q_i = -s_i + \frac{1}{2} r_i.
	\]
	Expanding their Hermitian product gives

	\begin{align*}
		\overline{(q_j - q_i)}(q_k - q_i)
		&= \left(-\bar{s}_i - \frac{1}{2}\bar{r}_i\right)\left(-s_i + \frac{1}{2} r_i\right) 
		= |s_i|^2 + \frac{1}{2}(\bar{r}_i s_i - \bar{s}_i r_i) - \frac{1}{4}|r_i|^2 \\[6pt]
		&= |s_i|^2 - \frac{1}{4}|r_i|^2 + \mathrm{i} \operatorname{Im}(\bar{r}_i s_i).
	\end{align*}
	Since $(i, j, k)$ is cyclic, the oriented double area is $B = \operatorname{Im}\bigl(\overline{(q_j - q_i)}(q_k - q_i)\bigr)$, so taking the imaginary part yields $B = \operatorname{Im}(\bar{r}_i s_i)$.
	
	Next, noting that $a_j = |q_k - q_i| = |s_i - r_i/2|$ and $a_k = |q_j - q_i| = |s_i + r_i/2|$, we expand $f_i = a_j^2 - a_k^2$ as
	\begin{align*}
		f_i
		&= \left|s_i - \frac{1}{2} r_i\right|^2 - \left|s_i + \frac{1}{2} r_i\right|^2 
		= \left(|s_i|^2 - \operatorname{Re}(\bar{r}_i s_i) + \frac{1}{4}|r_i|^2\right) - \left(|s_i|^2 + \operatorname{Re}(\bar{r}_i s_i) + \frac{1}{4}|r_i|^2\right) \\[6pt]
		&= -2\operatorname{Re}(\bar{r}_i s_i),
	\end{align*}
	which proves~\eqref{eq:B_fi_re_im}. Combining both parts of~\eqref{eq:B_fi_re_im} via $\bar{r}_i s_i = \operatorname{Re}(\bar{r}_i s_i) + \mathrm{i}\operatorname{Im}(\bar{r}_i s_i)$ immediately gives~\eqref{eq:ri_si_product}.
	
	Finally, \eqref{eq:Im_side_frame} is a well-known result for Jacobi coordinates, but can also be checked by hand similar to the above calculation.

\end{proof}

For equal masses, $(r_i, s_i)$ are the \textsc{Jacobi} vectors of
Definition~\ref{def:jacobi} for the pair $(j,k)$ and body~$i$, with reduced masses
$\mu = \tfrac12$ and $\nu = \tfrac23$; for $i = 3$ they coincide with $(r, z)$. In
\textsc{Jacobi} coordinates the moment of inertia and the angular momentum split without
mixed terms into a part of the pair and a part of body~$i$ relative to the midpoint of
the pair. For $I_m$ this is~\eqref{eq:Im_side_frame}; for $L$ it is the following lemma,
in which $\tfrac12 K_i = \mu\operatorname{Im}(\bar r_i\dot r_i)$ is the internal angular
momentum of the pair. Expressing $r_i$ in polar coordinates further makes this
rotational character explicit.

\begin{lemma}[Angular momentum decomposition and polar form of $K_i$]\label{lem:angular_momentum_Ki}
	For any cyclic permutation $(i, j, k)$ of $(1, 2, 3)$:
	\begin{enumerate}
		\item The total angular momentum $L$ decomposes in the Jacobi coordinates as
		\begin{equation}\label{eq:L_side_frame}
			L = \frac{2}{3}\operatorname{Im}(\bar{s}_i \dot{s}_i) + \frac{1}{2} K_i.
		\end{equation}
		In particular, along any zero-angular-momentum solution ($L = 0$), one has
		\begin{equation}\label{eq:Im_si_Ki}
			\operatorname{Im}(\bar{s}_i \dot{s}_i) = -\frac{3}{4} K_i.
		\end{equation}
		\item Whenever $a_i = |r_i| > 0$, writing $r_i = a_i e^{\mathrm{i}\varphi_i}$ with a local smooth angle $\varphi_i$ yields
		\begin{equation}\label{eq:Ki_polar}
			K_i = a_i^2 \dot{\varphi}_i.
		\end{equation}
	\end{enumerate}
\end{lemma}

\begin{proof}

\eqref{eq:L_side_frame} is a well-known statement from the theory of Jacobi coordinates, compare e.g. \cite{lim1991binary}, but it can also be checked by a short calculation. \eqref{eq:Im_si_Ki} follows immediately when $L = 0$.
	
For the second assertion, differentiate $r_i = a_i e^{\mathrm{i}\varphi_i}$ to obtain $\dot{r}_i = (\dot{a}_i + \mathrm{i} a_i \dot{\varphi}_i)e^{\mathrm{i}\varphi_i}$. Multiplying by $\bar{r}_i = a_i e^{-\mathrm{i}\varphi_i}$ gives
	\[
	\bar{r}_i \dot{r}_i = a_i e^{-\mathrm{i}\varphi_i} (\dot{a}_i + \mathrm{i} a_i \dot{\varphi}_i) e^{\mathrm{i}\varphi_i} = a_i \dot{a}_i + \mathrm{i} a_i^2 \dot{\varphi}_i,
	\]
	and taking the imaginary part yields $K_i = \operatorname{Im}(\bar{r}_i \dot{r}_i) = a_i^2 \dot{\varphi}_i$.
\end{proof}

\begin{remark}\label{rem:Ki_physical}
	The identification of $\tfrac12 K_i$ as internal angular momentum can also be checked
	directly: bodies~$j$ and~$k$ have unit masses, relative positions $\pm r_i/2$ and
	velocities $\pm \dot r_i/2$ with respect to their midpoint, so their angular momentum
	about that midpoint is
	\[
	\operatorname{Im}\!\left(\frac{\bar{r}_i}{2}\frac{\dot{r}_i}{2}\right)
	+ \operatorname{Im}\!\left(\left(-\frac{\bar{r}_i}{2}\right)\left(-\frac{\dot{r}_i}{2}\right)\right)
	= \frac{1}{2} K_i.
	\]
	By~\eqref{eq:Ki_polar}, $K_i $ thus measures the oriented rotation
	of the side opposite body~$i$.
\end{remark}

\subsection{Pair rotation dynamics and the torque equation}\label{subsec:rot_dynamics}

Body~$i$ attracts the closer endpoint of the opposite side more strongly and thereby turns this side. To make this precise, we express the acceleration of each side vector $r_i$ in terms of the coordinates $(r_i, s_i)$ using the equations of motion~\eqref{eq:alpha_forces}. This yields an evolution equation for $K_i$ whose sign, as long as $B > 0$, is opposite to the sign of $f_i$.

\begin{lemma}[Acceleration of the side vector]\label{lem:side_acceleration}
	Let $(i, j, k)$ be a cyclic permutation of $(1, 2, 3)$. Along any collision-free solution of~\eqref{eq:alpha_forces}, the side vector $r_i = q_k - q_j$ satisfies
	\begin{equation}\label{eq:ri_ddot}
		\ddot{r}_i = -\left(\frac{2}{a_i^{\alpha+2}} + \frac{1}{2a_j^{\alpha+2}} + \frac{1}{2a_k^{\alpha+2}}\right) r_i + \left(\frac{1}{a_j^{\alpha+2}} - \frac{1}{a_k^{\alpha+2}}\right) s_i.
	\end{equation}
\end{lemma}

\begin{proof}
	Recall that the side lengths are $a_i = |q_k - q_j|$, $a_j = |q_i - q_k|$, and $a_k = |q_i - q_j|$. Evaluating the equations of motion~\eqref{eq:alpha_forces} for bodies~$k$ and~$j$ gives
	\begin{align*}
		\ddot{q}_k &= \frac{q_j - q_k}{|q_j - q_k|^{\alpha+2}} + \frac{q_i - q_k}{|q_i - q_k|^{\alpha+2}} = \frac{q_j - q_k}{a_i^{\alpha+2}} + \frac{q_i - q_k}{a_j^{\alpha+2}}, \\[6pt]
		\ddot{q}_j &= \frac{q_k - q_j}{|q_k - q_j|^{\alpha+2}} + \frac{q_i - q_j}{|q_i - q_j|^{\alpha+2}} = \frac{q_k - q_j}{a_i^{\alpha+2}} + \frac{q_i - q_j}{a_k^{\alpha+2}}.
	\end{align*}
	By definition, $q_k - q_j = r_i$ and $q_j - q_k = -r_i$, while Lemma~\ref{lem:side_frame_inverse} expresses the remaining relative position vectors as
	\[
	q_i - q_j = s_i + \frac{1}{2} r_i, \qquad q_i - q_k = s_i - \frac{1}{2} r_i.
	\]
	Substituting these relations into the expressions for $\ddot{q}_k$ and $\ddot{q}_j$, we obtain
	\begin{align*}
		\ddot{q}_k &= -\frac{r_i}{a_i^{\alpha+2}} + \frac{1}{a_j^{\alpha+2}}\left(s_i - \frac{1}{2} r_i\right), \\[6pt]
		\ddot{q}_j &= \frac{r_i}{a_i^{\alpha+2}} + \frac{1}{a_k^{\alpha+2}}\left(s_i + \frac{1}{2} r_i\right).
	\end{align*}
	Subtracting the second equation from the first and collecting the terms proportional to $r_i$ and $s_i$ yields
	\begin{align*}
		\ddot{r}_i = \ddot{q}_k - \ddot{q}_j
		&= -\frac{2}{a_i^{\alpha+2}} r_i + \frac{1}{a_j^{\alpha+2}}\left(s_i - \frac{1}{2} r_i\right) - \frac{1}{a_k^{\alpha+2}}\left(s_i + \frac{1}{2} r_i\right) \\[6pt]
		&= -\left(\frac{2}{a_i^{\alpha+2}} + \frac{1}{2a_j^{\alpha+2}} + \frac{1}{2a_k^{\alpha+2}}\right) r_i + \left(\frac{1}{a_j^{\alpha+2}} - \frac{1}{a_k^{\alpha+2}}\right) s_i. \qedhere
	\end{align*}
\end{proof}

\begin{proposition}[Torque equation]\label{prop:torque_equation}
	For any cyclic permutation $(i, j, k)$ of $(1, 2, 3)$, the time derivative of $K_i = \operatorname{Im}(\bar{r}_i \dot{r}_i)$ is given by
	\begin{equation}\label{eq:Ki_dot}
		\dot{K}_i = B\left(a_j^{-(\alpha+2)} - a_k^{-(\alpha+2)}\right).
	\end{equation}
\end{proposition}

\begin{proof}
	Differentiating $K_i = \operatorname{Im}(\bar{r}_i \dot{r}_i)$ with respect to time gives
	\[
	\dot{K}_i = \operatorname{Im}\!\bigl(\dot{\bar{r}}_i \dot{r}_i + \bar{r}_i \ddot{r}_i\bigr) = \operatorname{Im}\!\bigl(|\dot{r}_i|^2\bigr) + \operatorname{Im}(\bar{r}_i \ddot{r}_i) = \operatorname{Im}(\bar{r}_i \ddot{r}_i),
	\]
	since $|\dot{r}_i|^2 \in \mathbb{R}$ has vanishing imaginary part. Substituting~\eqref{eq:ri_ddot} from Lemma~\ref{lem:side_acceleration} and using $\bar{r}_i r_i = |r_i|^2 \in \mathbb{R}$, we find
	\begin{align*}
		\dot{K}_i
		&= \operatorname{Im}\!\left[-\left(\frac{2}{a_i^{\alpha+2}} + \frac{1}{2a_j^{\alpha+2}} + \frac{1}{2a_k^{\alpha+2}}\right) |r_i|^2 + \left(a_j^{-(\alpha+2)} - a_k^{-(\alpha+2)}\right) \bar{r}_i s_i\right] \\[6pt]
		&= \left(a_j^{-(\alpha+2)} - a_k^{-(\alpha+2)}\right) \operatorname{Im}(\bar{r}_i s_i).
	\end{align*}
	Recalling from Proposition~\ref{prop:side_frame_geometry} that $\operatorname{Im}(\bar{r}_i s_i) = B$ yields~\eqref{eq:Ki_dot}.
\end{proof}

\begin{remark}\label{rem:torque_physical}
	In~\eqref{eq:ri_ddot}, every force component parallel to $r_i$—in particular the direct mutual attraction $-2r_i/a_i^{\alpha+2}$ between bodies~$j$ and~$k$—changes the side length $a_i = |r_i|$ but produces no torque on the pair, because $\operatorname{Im}(\bar{r}_i r_i) = 0$. The pair torque~\eqref{eq:Ki_dot} arises exclusively from the asymmetric attractive pull of the third body~$i$.
\end{remark}
\subsection{Kinematic relation between side asymmetry and area}\label{subsec:kinematic_relation}
The torque equation expresses the evolution of $K_i$ in terms of $f_i$. To close the loop, we need the converse: an equation in which $K_i$ drives the evolution of $f_i$. By~\eqref{eq:ri_si_product}, $f_i$ and $B$ are, up to a factor, the real and imaginary parts of the single complex quantity $\bar r_i s_i$, so we compare their evolution directly. The resulting identity is used both on $[0, T_*)$ and at the syzygy itself.

\begin{proposition}\label{prop:kinematic_relation}
	Along every zero-angular-momentum solution ($L=0$) of the equal-mass planar
	three-body problem and for each $i\in\{1,2,3\}$,
	\begin{equation}\label{eq:kinematic_identity}
		B\dot{f}_i - f_i\dot{B} = -3 I_m K_i .
	\end{equation}
\end{proposition}

\begin{proof}
	Set $Z_i := \bar r_i s_i$; we compute $\operatorname{Im}(\bar Z_i\dot Z_i)$ in two ways.
	By~\eqref{eq:ri_si_product}, $Z_i = -\tfrac12 f_i + \mathrm{i}B$, hence
	\begin{equation}\label{eq:Zi_im_part1}
		\operatorname{Im}(\bar Z_i\dot Z_i)
		= \operatorname{Im}\Bigl[\bigl(-\tfrac12 f_i - \mathrm{i}B\bigr)\bigl(-\tfrac12\dot f_i + \mathrm{i}\dot B\bigr)\Bigr]
		= \tfrac12\bigl(B\dot f_i - f_i\dot B\bigr).
	\end{equation}
	On the other hand, $\bar Z_i = r_i\bar s_i$ and $\dot Z_i = \dot{\bar r}_i s_i + \bar r_i\dot s_i$ give
	\[
	\bar Z_i\dot Z_i = |s_i|^2\, r_i\dot{\bar r}_i + |r_i|^2\, \bar s_i\dot s_i .
	\]
	Since $\operatorname{Im}(r_i\dot{\bar r}_i) = -\operatorname{Im}(\bar r_i\dot r_i) = -K_i$ and,
	by~\eqref{eq:Im_si_Ki} we have
	$\operatorname{Im}(\bar s_i\dot s_i) = -\tfrac34 K_i$ (the only place where $L=0$ enters), we obtain with~\eqref{eq:Im_side_frame}
	\begin{equation}\label{eq:Zi_im_part2}
		\operatorname{Im}(\bar Z_i\dot Z_i)
		= -\Bigl(|s_i|^2 + \tfrac34|r_i|^2\Bigr)K_i
		= -\tfrac32 I_m K_i .
	\end{equation}
	Comparing~\eqref{eq:Zi_im_part1} and~\eqref{eq:Zi_im_part2} proves the claim.
\end{proof}

\begin{remark}\label{rem:kinematic_global}
	Wherever $B\neq 0$, \eqref{eq:kinematic_identity} is equivalent to
	\[
	\frac{d}{dt}\Bigl(\frac{f_i}{B}\Bigr) = -\frac{3 I_m K_i}{B^2}.
	\]
	The form~\eqref{eq:kinematic_identity}, however, is polynomial in the components
	of $r_i, s_i$ and their derivatives, and so are the identities
	\eqref{eq:ri_si_product} and \eqref{eq:Im_si_Ki} it rests on. It therefore holds
	at every time at which the solution is smooth, in particular at syzygies,
	where $B=0$ and the quotient form breaks down.
\end{remark}

\subsection{Preservation of the side ordering on \texorpdfstring{$[0,T_*)$}{[0,T*)}}\label{subsec:ordering_open}

In this section we assume $\alpha>0$. The condition $\alpha<2$ will be only necessary in the following subsections. Recall that brake orbits have $L = 0$ by Lemma~\ref{lem:conservation}, so Proposition~\ref{prop:kinematic_relation} applies. The mechanism is a feedback loop: $K_i$ vanishes at $t = 0$, the torque equation~\eqref{eq:Ki_dot} forces $K_i$ to take the sign opposite to $f_i$, and by Remark~\ref{rem:kinematic_global} this in turn makes $f_i/B$ move away from zero.

\begin{proposition}\label{prop:side_comparison_preservation}
	Let $q$ be an equal-mass brake solution with $B(0) > 0$, and let $i \in \{1,2,3\}$ with $f_i(0) \neq 0$. Then
	\[
	\operatorname{sgn} f_i(t) = \operatorname{sgn} f_i(0) \quad (0 \le t < T_*),
	\qquad
	\operatorname{sgn} K_i(t) = -\operatorname{sgn} f_i(0) \quad (0 < t < T_*),
	\]
	and $|K_i|$ is strictly increasing on $[0, T_*)$.
\end{proposition}

\begin{proof}
	We treat the case $f_i(0) > 0$; the case $f_i(0) < 0$ is identical with all inequalities reversed. Since $\dot q(0) = 0$, we have $\dot r_i(0) = 0$ and hence $K_i(0) = 0$. Let
	\[
	\tau := \sup\bigl\{ t \in [0, T_*) \,\big|\, f_i > 0 \text{ on } [0, t] \bigr\},
	\]
	so that $\tau > 0$ by continuity. On $[0, \tau)$ we have $a_j > a_k$, hence $a_j^{-(\alpha+2)} < a_k^{-(\alpha+2)}$, and $B > 0$; the torque equation~\eqref{eq:Ki_dot} gives $\dot K_i < 0$. Thus $K_i$ is strictly decreasing on $[0,\tau)$ with $K_i < 0$ on $(0, \tau)$. Since $B > 0$ and $I_m > 0$ on $[0, T_*)$, Remark~\ref{rem:kinematic_global} yields
	\[
	\frac{d}{dt}\Bigl(\frac{f_i}{B}\Bigr) = -\frac{3 I_m K_i}{B^2} > 0 \qquad (0 < t < \tau),
	\]
	and therefore $f_i(t)/B(t) \ge f_i(0)/B(0) > 0$ for $t \in [0, \tau)$. If $\tau < T_*$, continuity gives $f_i(\tau) \ge B(\tau) f_i(0)/B(0) > 0$, so $f_i > 0$ on $[0, \tau + \eta]$ for some $\eta > 0$, contradicting the definition of $\tau$. Hence $\tau = T_*$.
\end{proof}

\begin{remark}[Isosceles data]\label{rem:isosceles}
	If $f_i(0) = 0$, then $f_i \equiv K_i \equiv 0$ on $[0, T_*)$: an initially isosceles triangle remains isosceles. To see this, apply the fundamental theorem of calculus to \\ $u \mapsto u^{-(\alpha+2)/2}$ between $a_k^2$ and $a_j^2$, which gives
	\[
	a_j^{-(\alpha+2)} - a_k^{-(\alpha+2)} = -g_i f_i,
	\qquad
	g_i := \frac{\alpha+2}{2}\int_0^1 \bigl((1-\xi)a_k^2 + \xi a_j^2\bigr)^{-(\alpha+4)/2}\,d\xi > 0.
	\]
	Unlike the difference quotient, $g_i$ is continuous also where $a_j = a_k$. By~\eqref{eq:Ki_dot} and Remark~\ref{rem:kinematic_global}, the pair $(f_i/B, K_i)$ satisfies
	\[
	\frac{d}{dt}\Bigl(\frac{f_i}{B}\Bigr) = -\frac{3 I_m}{B^2}\, K_i,
	\qquad
	\dot K_i = -B^2 g_i\, \frac{f_i}{B}
	\qquad (0 \le t < T_*),
	\]
	a homogeneous linear system whose coefficients are continuous along the given solution. Since both components vanish at $t = 0$, uniqueness gives $f_i/B \equiv K_i \equiv 0$ on $[0, T_*)$.
\end{remark}

\begin{corollary}\label{cor:strict_side_ordering}
	Let $q$ be an equal-mass brake solution with $B(0) > 0$ and $a_1(0) > a_2(0) > a_3(0)$. Then
	\[
	a_1(t) > a_2(t) > a_3(t) > 0 \qquad (0 \le t < T_*),
	\]
	and $K_1, K_3 < 0 < K_2$ on $(0, T_*)$, with $|K_1|, |K_2|, |K_3|$ strictly increasing on $[0, T_*)$.
\end{corollary}

\begin{proof}
	We have $f_1(0) = a_2(0)^2 - a_3(0)^2 > 0$, $f_2(0) = a_3(0)^2 - a_1(0)^2 < 0$ and $f_3(0) = a_1(0)^2 - a_2(0)^2 > 0$. Proposition~\ref{prop:side_comparison_preservation} gives the signs of $f_i$ and $K_i$; $f_3 > 0$ and $f_1 > 0$ mean $a_1 > a_2 > a_3$, and $a_3 > 0$ because $q(t) \notin \Delta$ on $[0, T_*)$.
\end{proof}

Corollary~\ref{cor:strict_side_ordering} gives the ordering only on the half-open interval $[0, T_*)$. Two things remain: we must exclude a collision at $T_*$, so that $T_*$ is a regular syzygy, and we must show that no two sides become equal at $T_*$. The assumption $\alpha < 2$ enters only through the first point, via Lemma~\ref{lem:distance_lower_bound}.

\subsection{A lower bound on the side lengths}\label{subsec:distance_lower_bound}

From now on we assume $0 < \alpha < 2$, $m_1 = m_2 = m_3 = 1$, $B(0) > 0$ and $a_1(0) > a_2(0) > a_3(0)$.

\begin{lemma}\label{lem:distance_lower_bound}

	For each initial configuration there exists $\delta > 0$ depending on the trajectory such that
	\begin{equation}\label{eq:distance_lower_bound}
		a_i(t) \ge \delta \qquad (i \in \{1,2,3\},\; 0 \le t < T_*).
	\end{equation}
\end{lemma}

\begin{proof}
	By Corollary~\ref{cor:strict_side_ordering}, $a_3$ is the shortest side on $[0, T_*)$, so it suffices to bound $a_3$. Fix $t_0 \in (0, T_*)$ and set $\gamma := -K_3(t_0)$. By Corollary~\ref{cor:strict_side_ordering}, $\gamma > 0$ and
	\begin{equation}\label{eq:K3_lower_bound}
		-K_3(t) \ge \gamma \qquad (t_0 \le t < T_*).
	\end{equation}
	For unit masses and $\alpha G = 1$, the potential is $U = \alpha^{-1}\sum_{i=1}^3 a_i^{-\alpha}$. Energy conservation~\eqref{eq:energy} gives $\mathcal T = U - U_0$, and since $a_3$ is the shortest side,
	\begin{equation}\label{eq:T_upper_bound}
		\mathcal T = U - U_0 \le U \le \frac{3}{\alpha a_3^\alpha}.
	\end{equation}
	Since $r_3 = q_2 - q_1$ and $|r_3| = a_3$, we obtain for $t \in [t_0, T_*)$
	\begin{equation}\label{eq:k2_a3_bound}
		\gamma^2 \le K_3^2 \le |r_3|^2 |\dot r_3|^2
		= a_3^2 |\dot q_2 - \dot q_1|^2
		\le 2 a_3^2 \bigl(|\dot q_1|^2 + |\dot q_2|^2\bigr)
		\le 4 a_3^2 \mathcal T
		\le \frac{12}{\alpha}\, a_3^{2-\alpha}.
	\end{equation}
	As $2 - \alpha > 0$, this gives $a_3(t) \ge (\alpha \gamma^2/12)^{1/(2-\alpha)}$ on $[t_0, T_*)$. On the compact interval $[0, t_0]$, the continuous function $a_3$ is positive and attains a positive minimum. Hence
	\[
	\delta := \min\Bigl\{ \min_{[0, t_0]} a_3,\; \bigl(\alpha \gamma^2/12\bigr)^{1/(2-\alpha)} \Bigr\} > 0
	\]
	satisfies~\eqref{eq:distance_lower_bound}.
\end{proof}

\begin{remark}\label{rem:energetic_obstruction}
	Estimate~\eqref{eq:k2_a3_bound} shows where $\alpha < 2$ enters: a pair rotation $|K_3| \ge \gamma$ forces $\mathcal T \ge \gamma^2/(4a_3^2)$, whereas by~\eqref{eq:T_upper_bound} at most $3/(\alpha a_3^\alpha)$ is available. For $\alpha < 2$, the first quantity outgrows the second as $a_3 \to 0$. So the angular momentum creates a barrier for collisions.
\end{remark}

\subsection{Continuation through the first syzygy}\label{subsec:continuation_syzygy}

With collisions excluded, the solution cannot break down at $T_*$; it must reach a collinear configuration.

\begin{proposition}\label{prop:regular_syzygy_continuation}
	The solution extends smoothly to $[0, T_* + \varepsilon)$ for some $\varepsilon > 0$, and
	\[
	B(T_*) = 0, \qquad a_i(T_*) \ge \delta \quad (i = 1, 2, 3).
	\]
	In particular, $T_*$ is the first regular syzygy time in the sense of Definition~\ref{def:Tstar}.
\end{proposition}

\begin{proof}
	By Lemma~\ref{lem:sqrtB_concave}, $T_* < \infty$. Let $[0, \omega)$ be the forward interval of existence of the maximal solution, so $\omega \ge T_*$. For $t \in [0, T_*)$ we have
	\begin{itemize}
		\item $|q_i(t)|^2 \le I_m(t) \le I_m(0)$ by Theorem~\ref{thm:inertia},
		\item $|q_j(t) - q_i(t)| \ge \delta$ by Lemma~\ref{lem:distance_lower_bound},
		\item $|\dot q_i(t)|^2 \le 2\mathcal T(t) \le 6/(\alpha\delta^\alpha)$ by~\eqref{eq:T_upper_bound} and $a_3 \ge \delta$.
	\end{itemize}
	Hence $(q, \dot q)$ stays in a compact subset of the phase space $\mathcal P$ on $[0, T_*)$. Since a maximal solution of an ODE with locally Lipschitz vector field leaves every compact subset of its domain as $t \to \omega$ whenever $\omega < \infty$, we conclude $\omega > T_*$. By continuity, $a_i(T_*) \ge \delta$ and $B(T_*) \ge 0$. If $B(T_*) > 0$, then $B > 0$ and $q \notin \Delta$ on $[0, T_* + \eta]$ for some $\eta > 0$, contradicting the definition of $T_*$. Hence $B(T_*) = 0$.
\end{proof}

\subsection{The ordering at \texorpdfstring{$T_*$}{T*}}\label{subsec:ordering_at_syzygy}

At $T_*$ we have $B = 0$, so the monotonicity of $f_i/B$ used in Proposition~\ref{prop:side_comparison_preservation} is no longer available. Instead we evaluate the kinematic identity~\eqref{eq:kinematic_identity} directly at $T_*$; this is where its division-free form is needed.

\begin{proposition}\label{prop:endpoint_ordering}
	If $f_i(0) \neq 0$, then $f_i(T_*) \neq 0$ has the same sign as $f_i(0)$. In particular,
	\begin{equation}\label{eq:closed_interval_ordering}
		a_1(t) > a_2(t) > a_3(t) > 0 \qquad (0 \le t \le T_*).
	\end{equation}
\end{proposition}

\begin{proof}
	By Proposition~\ref{prop:regular_syzygy_continuation}, all quantities involved are smooth on $[0, T_*]$, $B(T_*) = 0$ and $I_m(T_*) > 0$. By Proposition~\ref{prop:side_comparison_preservation}, $f_i$ has the sign of $f_i(0)$ on $[0, T_*)$, so $f_i(T_*)$ has the same sign or vanishes. Suppose $f_i(T_*) = 0$. By Remark~\ref{rem:kinematic_global}, \eqref{eq:kinematic_identity} holds at $t = T_*$, and its left-hand side vanishes there. Hence $I_m(T_*) K_i(T_*) = 0$, i.e.\ $K_i(T_*) = 0$. On the other hand, $|K_i|$ is strictly increasing on $[0, T_*)$ by Proposition~\ref{prop:side_comparison_preservation}, so for any $s \in (0, T_*)$,
	\[
	|K_i(T_*)| = \lim_{t \nearrow T_*} |K_i(t)| \ge |K_i(s)| > 0,
	\]
	a contradiction. Applying this to $f_1(T_*) > 0$ and $f_3(T_*) > 0$ gives $a_1(T_*) > a_2(T_*) > a_3(T_*)$; together with Corollary~\ref{cor:strict_side_ordering} and $a_3(T_*) \ge \delta$, this proves~\eqref{eq:closed_interval_ordering}.
\end{proof}

\begin{proof}[Proof of Theorem~\ref{thm:ordering}]
	After relabeling the bodies and, if necessary, reflecting the plane (which does not change side lengths), we may assume $a_1(0) > a_2(0) > a_3(0)$ and $B(0) > 0$. By Lemma~\ref{lem:sqrtB_concave} and Proposition~\ref{prop:regular_syzygy_continuation}, $T_*$ is a finite first regular syzygy time and the solution is collision-free on $[0, T_*]$. Proposition~\ref{prop:endpoint_ordering} gives~\eqref{eq:orderstatement}.
\end{proof}

\begin{remark}[Middle body]\label{rem:middle_body}
	At the collinear configuration $q(T_*)$, the three bodies are pairwise distinct, so the longest side is the sum of the other two and joins the two outer bodies. By~\eqref{eq:closed_interval_ordering}, this side is $a_1 = r_{23}$, so body~$1$ lies between bodies~$2$ and~$3$. Thus the middle body at the first syzygy is the body opposite the initially longest side, equivalently the vertex with the largest initial interior angle.
\end{remark}

\subsection{An explicit bound on \texorpdfstring{$T_*$}{T*}}\label{subsec:finite_time_angle}

We close with a quantitative complement to Lemma~\ref{lem:sqrtB_concave}, which already shows that $T_*$ is finite. The interior angle $\theta_1(t) \in (0, \pi)$ at body~$1$, i.e.\ the largest angle of the triangle, yields an explicit bound. The argument does not use $\alpha < 2$.

\begin{corollary}\label{cor:finite_time_angle}
	For every $t_0 \in (0, T_*)$,
	\begin{equation}\label{eq:Tstar_angle_bound}
		T_* \le t_0 + \frac{\pi - \theta_1(t_0)}{c_0},
		\qquad
		c_0 := -K_3(t_0)\left(\frac{\alpha U_0}{3}\right)^{2/\alpha} > 0.
	\end{equation}
\end{corollary}

\begin{proof}
	Since $\mathcal T \ge 0$, \eqref{eq:T_upper_bound} gives $U_0 \le 3/(\alpha a_3^\alpha)$, i.e.
	\begin{equation}\label{eq:a3_upper_bound}
		a_3(t) \le \left(\frac{3}{\alpha U_0}\right)^{1/\alpha} \qquad (0 \le t < T_*).
	\end{equation}
	The sides at body~$1$ are $r_3 = q_2 - q_1$ and $-r_2 = q_3 - q_1$. Since $B = \operatorname{Im}\bigl(\bar r_3(-r_2)\bigr) > 0$ on $[0, T_*)$, $\theta_1$ is the counterclockwise angle from $r_3$ to $-r_2$. With local smooth polar angles $\varphi_2, \varphi_3$ of $r_2, r_3$, we have $\theta_1 = \varphi_2 + \pi - \varphi_3$ modulo $2\pi$, and~\eqref{eq:Ki_polar} gives
	\[
	\dot\theta_1 = \dot\varphi_2 - \dot\varphi_3 = \frac{K_2}{a_2^2} - \frac{K_3}{a_3^2}.
	\]
	For $t \in [t_0, T_*)$, Corollary~\ref{cor:strict_side_ordering} gives $K_2 > 0$ and $-K_3(t) \ge -K_3(t_0)$; together with~\eqref{eq:a3_upper_bound},
	\begin{equation}\label{eq:theta1_dot}
		\dot\theta_1(t) > \frac{-K_3(t)}{a_3(t)^2} \ge \frac{-K_3(t_0)}{a_3(t)^2} \ge c_0.
	\end{equation}
	Integrating yields $\pi > \theta_1(t) \ge \theta_1(t_0) + c_0(t - t_0)$ for all $t \in [t_0, T_*)$, and letting $t \to T_*$ proves~\eqref{eq:Tstar_angle_bound}.
\end{proof}

\section{Outlook}
\label{sc:outlook}

We conclude with remarks on the scope of our methods and on open questions.

\subsection{Other exponents}

Under the normalization $\alpha G = 1$, the equations of motion
\[
\ddot q_i = \sum_{j \neq i} m_j \frac{q_j - q_i}{|q_j - q_i|^{\alpha+2}}
\]
describe attractive central forces for every $\alpha \in \mathbb{R}$. For $\alpha < 0$ this corresponds to $G < 0$, and $\alpha = 0$ to the logarithmic potential $U = -\sum_{i<j} m_i m_j \log r_{ij}$. It is natural to ask which parts of our results survive in this wider range.

\emph{Theorem~\ref{thm:inertia}.} The proof in Section~\ref{sc:theorem_inertia} uses only that the coefficients $m_i m_j r_{ij}^{-(\alpha+2)}$ in $\mathcal S$ are positive; neither the sign of $\alpha$ nor the sign of the energy enters. Theorem~\ref{thm:inertia} therefore holds for every $\alpha \in \mathbb{R}$.

\emph{Theorem~\ref{thm:ordering} for $-2 < \alpha \le 0$.} The sign preservation of Section~\ref{subsec:ordering_open} uses only that $x \mapsto x^{-(\alpha+2)}$ is strictly decreasing, i.e.\ $\alpha > -2$. The assumption $\alpha > 0$ enters the rest of Section~\ref{sc:thm_ordering} only through the energy estimate~\eqref{eq:T_upper_bound}. For $-2 < \alpha < 0$ one has $U \le 0$, hence $\mathcal T \le |U_0|$, and~\eqref{eq:k2_a3_bound} becomes $\gamma^2 \le 4|U_0|\,a_3^2$. For $\alpha = 0$ one has $U \le -3\log a_3$, and~\eqref{eq:k2_a3_bound} becomes $\gamma^2 \le 4 a_3^2\bigl(|U_0| + 3|\log a_3|\bigr)$. In both cases $a_3$ stays bounded away from zero, which suggests that Theorem~\ref{thm:ordering} extends to $-2 < \alpha < 2$ with essentially the same proof.

\emph{The harmonic case $\alpha = -2$.} Here $\ddot q_i = -3 q_i$ for equal masses, so every brake orbit is homothetic, $q_i(t) = q_i(0)\cos(\sqrt 3\, t)$, and ends in a triple collision at $t = \pi/(2\sqrt 3)$. The torque~\eqref{eq:Ki_dot} vanishes identically. The ordering is trivially preserved, but the first syzygy is not regular, so Theorem~\ref{thm:ordering} fails at $\alpha = -2$. For $\alpha < -2$ the torque has the same sign as $f_i$, the feedback between $K_i$ and $f_i/B$ becomes restoring rather than self-reinforcing, and we do not expect the ordering to be preserved in general.

\emph{The range $\alpha\geq2$.}
The collision-exclusion argument of Lemma~\ref{lem:distance_lower_bound} relies on $2-\alpha>0$ and therefore does not cover $\alpha \ge 2$. Corollary~\ref{cor:strict_side_ordering} still gives the ordering on $[0,T_*)$, and Proposition~\ref{prop:endpoint_ordering} applies whenever $T_*$ is a regular syzygy.

The borderline case $\alpha=2$ admits a separate
treatment in our Lean formalization, but lies
outside the scope of the present paper.
For $\alpha>2$, the attraction $r^{-\alpha}$
is more singular than the centrifugal barrier
of order $r^{-2}$, so the collision-exclusion
argument requires further investigation.
The case $\alpha=2$ is special for another reason:
the \textsc{Lagrange}--\textsc{Jacobi} identity
$\ddot I_m = 4E + (4 - 2\alpha)U$ reduces to
$\ddot I_m = -4U_0$, so
$I_m(t) = I_m(0) - 2U_0 t^2$ is explicit.

\subsection{More bodies and higher dimensions}

Three bodies released from rest remain in the plane spanned by their initial positions, since all forces lie in this plane and the initial velocities vanish. Our results therefore hold verbatim in $\mathbb{R}^d$ for every $d \ge 2$.

A genuine higher-dimensional analogue concerns $n$ bodies in $\mathbb{R}^{n-1}$, released from rest at the vertices of a nondegenerate simplex; the analogue of a syzygy is the first time the bodies lie in a common hyperplane. The configuration matrix $Q$ is then a square $(n-1)\times(n-1)$ matrix whose determinant is proportional to the oriented volume of the simplex, and $\mathcal S$ remains positive definite. The argument of Section~\ref{sc:theorem_inertia} appears to carry over: $C = Q^{-1}\dot Q$ is symmetric and negative definite, $I_m$ decreases strictly, and with $m = n-1$ the inequality $(\operatorname{tr} C)^2 \le m \operatorname{tr}(C^2)$ yields
\[
\frac{1}{(\det Q)^{1/m}} \frac{d^2}{dt^2} (\det Q)^{1/m}
= \frac1m\Bigl(\frac1m(\operatorname{tr} C)^2 - \operatorname{tr}(C^2) - \operatorname{tr}\mathcal S\Bigr) < 0,
\]
so the simplex degenerates in finite time. For $n \ge 4$ bodies in the plane, $Q$ is no longer square and the method does not apply directly. An analogue of Theorem~\ref{thm:ordering} for more bodies, and the case of unequal masses for three bodies, remain open.

\subsection{General pair forces}

Theorem~\ref{thm:inertia} already holds for the general attractive central pair forces \eqref{eq:general_central_forces}. For equal unit masses and a common pair coefficient $\kappa_{ij}(r)=\varphi(r)$, the torque equation becomes $\dot K_i = B\bigl(\varphi(a_j)-\varphi(a_k)\bigr)$. Thus, the argument for strict side-order preservation on $[0,T_*)$ extends whenever $\varphi$ is strictly decreasing. To obtain a regular first syzygy, however, one must additionally exclude collisions. The estimate of Lemma~\ref{lem:distance_lower_bound} suggests that the singularity of the associated pair potential, in comparison with $r^{-2}$, plays an important role, but we do not prove a general criterion here.

\subsection{Physical context}

For $\alpha = d - 2$, the $\alpha$-potential is the fundamental solution of the \textsc{Laplace} equation in $\mathbb{R}^d$, i.e.\ Newtonian gravity in $d$ spatial dimensions. Since the motion of three bodies released from rest is planar, our results describe free fall of three bodies under the gravity of a $d$-dimensional world: Theorem~\ref{thm:inertia} for every $d$, Theorem~\ref{thm:ordering} for $d = 3$ and, by the discussion above, presumably for $d = 2$, while $d = 4$ is exactly the borderline case $\alpha = 2$.

In astronomy, the free-fall three-body problem is a classical model for the dynamical evolution of small stellar groups formed at rest. It goes back to \textsc{Burrau}'s Pythagorean problem~\cite{Burrau1913,SzebehelyPeters1967}, with masses $3, 4, 5$ released from rest at the vertices of a $3$-$4$-$5$ triangle, and was studied systematically by \textsc{Agekyan} and \textsc{Anosova}~\cite{AgekyanAnosova1967}; for recent work on free-fall periodic orbits see~\cite{HHDT2024}. Theorem~\ref{thm:inertia} applies to unequal masses and hence, for instance, to the Pythagorean problem. Theorem~\ref{thm:ordering} gives a rigorous statement about the early phase of the equal-mass free-fall evolution: the shape of the initial triangle determines which body is in the middle at the first syzygy.

\section*{Disclosure of AI use}

This paper was written as a part of the Principia Fall competition \url{https://principia-math.com/projects/6fd385f995b49c7a4d929f6f47b979a9/overview}, without which this project would not have been realized like this. The main results were obtained by GPT-6 Astra prompted by M.Q. in a discussion about approaches for the initial questions posed in \cite{MMV}. The other authors then explored the result as part of the challenge, whereas M.Q. was helping as a supervisor. There are also some other parts of this challenge (Lean formalization, exposition, writing of the paper) were different AI systems were used supporting the authors. We have checked all calculations by ourselves and improved on the original AI output.

\section*{Acknowledgement}

M.Q. wants to thank Nathan Duignan, who told him about these open questions in 2024 and discussed some initial approaches. We also want to thank Richard Montgomery for posing the questions originally and his interest in our work.

\section*{Notes in proof}

We have realized on November 28th, that the problems solved in this paper were included in a database of open problems \url{https://lefttosolve.com/p/mo-13933}. On November 30th, there was a Lean solution of these statements. We want to emphasize that we obtained our results independent from this formalization and the work of other researchers.

\bibliographystyle{alpha}
\bibliography{literatur}

\end{document}